\documentclass[a4paper,12pt]{article}
\usepackage[left=3.17cm,right=3.17cm,top=2.54cm,
headheight=0.5cm,headsep=0.54cm,bottom=2.54cm,footskip=0.79cm
]{geometry}

\usepackage{amsmath,amsthm,amssymb,cite}

\usepackage{hyperref}

\newtheorem{corollary}{Corollary}
\newtheorem{definition}{Definition}
\newtheorem{lemma}{Lemma}
\newtheorem{proposition}{Proposition}
\newtheorem{theorem}{Theorem}

\begin{document}

\title{Connes spectral distance as a quantum distance measure and a geometric measure of entanglement}

\author{Bing-Sheng Lin$^{1,2,\dag}$, Ji-Hong Wang$^{3,\ddag}$\\
	\small $^{1}$ School of Mathematics, South China University of Technology,
	Guangzhou 510641, China\\
	\small $^{2}$ Institute for Mathematics, Astrophysics and Particle Physics,\\
	\small  Radboud University Nijmegen, Heyendaalseweg 135, 6525 AJ Nijmegen, The Netherlands\\
	\small $^{\dag}$ E-mail address: sclbs@scut.edu.cn\\
	\small $^{3}$ Guangzhou University of Software, Guangzhou 510990, China\\
	\small $^{\ddag}$ E-mail address: wih@mail.seig.edu.cn
}

\date{\today}

\maketitle

\begin{abstract}
	In this paper, we study the Connes spectral distance as a quantum distance measure and its applications in discriminating quantum states and calculating quantum entanglement. Based on finite matrix algebra, we construct a spectral triple for finite dimensional quantum states. The Connes spectral distance is a generalization to the quantum trace distance and Hilbert-Schmidt distance. The corresponding optimal elements can always be chosen to commute with the difference matrix $\Delta$, and then the optimal element and $\Delta$ can be simultaneously diagonalized.
	We obtain the formulas of Connes spectral distances in three different parameter regimes which depending on the spectrum of difference matrix.
	As applications, we show how to use spectral distances to distinguish state pairs that have the same trace distance and Hilbert-Schmidt distance. We also define a new geometric measure of entanglement based on spectral distance.
	These results can help us study the relationships among Connes spectral distance, quantum trace distance and other quantum distance measures. These concrete constructions are significant for studies of quantum states in the framework of noncommutative geometry.\\
	
	
	
\end{abstract}

\section{Introduction}

In the framework of noncommutative geometry \cite{Connes1994},
a noncommutative space can be described by a spectral triple. Due to the noncommutativity, there is no traditional points in noncommutative spaces.
Pure states of a noncommutative space play the role of points in classical commutative geometry,
and traditional distances between points are represented by Connes spectral distances between states \cite{Connes1}. Since
the Connes spectral distance is a distance measure of quantum states, it may characterize the relationships and also intrinsic properties of the states.
In the past decades, the Connes spectral distances in different kinds of noncommutative spaces have already been studied in literature
\cite{Bimonte,Dai,Cagnache,Wallet,Martinetti,DAndrea,Pythagoras,Pythagoras1,Scholtz,Chaoba,Martinettip,Revisiting,Kumar,Chakraborty,Lin1,Lin2,Clare,Kumar1,LX,Lin4}.

In quantum mechanics,
most physical properties are invariant under unitary transformations of states.
Therefore,
to describe quantum systems and quantum states using spectral triples and Connes spectral distances,
one can also require Connes spectral distances to be unitarily invariant \cite{Lin3}.
In quantum information theory, there are many types of distance measures, such as
the quantum trace distance and Hilbert--Schmidt distance \cite{Nielsen},
both of which are naturally unitarily invariant. Therefore,
it is reasonable to study some types of Connes spectral distance as the generalization of conventional quantum trace and Hilbert-Schmidt distances.
Similar to normal quantum distance measures,
one can also use these types of spectral distances to investigate relations between quantum states and also properties of states, such as quantum entanglement.

This paper is organized as follows. In Sec.~\ref{sec2}, we first review the definitions of spectral triples and Connes spectral distances. We construct a spectral triple for the quantum states, which also has a real structure and is even with KO-dimension 6. We construct three types of Dirac operator corresponding to the quantum trace distance, Hilbert-Schmidt distance and a kind of generalized spectral distance, respectively.
In Sec.~\ref{sec3}, we study Connes spectral distances and optimal elements. We find that the optimal elements can always be chosen to commute with the difference matrix $\Delta$, and the optimal element and $\Delta$ can be simultaneously diagonalized.
In Sec.~\ref{sec4}, we study the formulas of Connes spectral distances in three different parameter regimes.
In Sec.~\ref{sec5}, we derive the exact thresholds of the parameter depending on the spectrum of difference matrix.
In Sec.~\ref{sec6}, we obtain the spectral distance function between arbitrary pairs of
pure states. We find that the result is similar to normal trace distance.
In Sec.~\ref{sec7}, we show how to use spectral distances to distinguish state pairs that have the same trace distance and Hilbert-Schmidt distance.
In Sec.~\ref{sec8}, we define a new geometric measure of entanglement based on spectral distance. We explicitly calculate the spectral geometric entanglement of Bell-diagonal states. We also compare the results with the
concurrence and negativity.
Some discussions and conclusions are given in Sec.~\ref{sec9}.

\section{Spectral Triples and Connes Spectral Distances}\label{sec2}

\begin{definition}
  \cite{Connes1994} A finite real spectral triple $(A, H, D, \Gamma, J)$ is defined as follows:
  \begin{itemize}
    \item A finite-dimensional Hilbert space $H$.
    
    \item A $\ast$-algebra $A$ acting faithfully from the left on $H$.
    
    \item An integer $d$ modulo $8$, called the KO-dimension.
    
    \item For even KO-dimension $d$,
    there is a Hermitian operator $\Gamma : H \rightarrow H$, called the chirality operator,
    which commutes with the algebra $A$, $a \Gamma = \Gamma a$, and satisfies $\Gamma^2 =
    1$.
    
    \item An antiunitary map $J : H \rightarrow H$, called the real structure, satisfying
    \begin{equation}
      [a, J b J^{- 1}] = 0, \quad \forall a, b \in A,
    \end{equation}
    together with $J^2 = \varepsilon$. For even KO-dimension $d$, one also has $J \Gamma =
    \varepsilon'' \Gamma J$.
    
    \item A Hermitian operator $D : H \rightarrow H$, called the Dirac operator,
    satisfying
    \begin{equation}
      [[D, a], J b J^{- 1}] = 0, \quad \forall a, b \in A,
    \end{equation}
    together with $D J = \varepsilon' J D$. For even KO-dimension $d$, there is $D \Gamma +
    \Gamma D = 0$.
  \end{itemize}
\end{definition}

The relationship between the signs $\varepsilon, \varepsilon',
\varepsilon''$ and the KO-dimension is as follows:
\[ {\begin{array}{|c|c|c|c|c|c|c|c|c|}
     \hline
     d & 0 & 1 & 2 & 3 & 4 & 5 & 6 & 7\\
     \hline
     \varepsilon & 1 & 1 & - 1 & - 1 & - 1 & - 1 & 1 & 1\\
     \hline
     \varepsilon' & 1 & - 1 & 1 & 1 & 1 & - 1 & 1 & 1\\
     \hline
     \varepsilon'' & 1 &  & - 1 &  & 1 &  & - 1 & \\
     \hline
   \end{array}}  \]

A state $\psi$ on the algebra $A$ is a linear functional $\psi : A \to
\mathbb{C}$ that is positive, i.e., $\psi(a^{\ast}a) \ge
0$ for $a\in A$, and has norm $1$. A normal bounded state
$\psi$ can also be represented by a density operator $\rho \in A$, where $\rho \geqslant
0$, $\mathrm{Tr} \rho = 1$. 

In this work, we mainly consider finite-dimensional quantum states, so we take $A =
\mathbb{M}_n (\mathbb{C})$. The action of the state $\psi$ on $a\in A$ can be written as
\begin{equation}
  \psi (a) = \mathrm{Tr} (\rho a) .
\end{equation}

Given a faithful representation $\pi : A \to \mathcal{B} (H)$ and a self-adjoint operator
$D$, define the map
\begin{equation}
  L_D (a) := \| [D, \pi (a)] \|_{op}, \label{eq:Connes}
\end{equation}
where $\|a\|_{o p}$ denotes the operator norm of the operator $a$:
\begin{equation}
  \|a\|_{o p} \equiv \sup_{\xi \in H, \hspace{0.17em} \| \xi \| = 1} \|a \xi
  \|, \qquad \|a\|^2 \equiv \mathrm{Tr} (a^{\dag} a) .
\end{equation}
The map $L_D(a)$ is a Lipschitz seminorm. Define the set
\[ B := \{a \in A : L_D (a) \leqslant 1\} . \]
The inequality $L_D (a) = \| [D, \pi (a)] \|_{o p} \leqslant 1$
is the so-called ball condition.

\begin{definition}
  \cite{Connes1} In a spectral triple $(A,H,D)$, let the quantum states $\psi_1,\psi_2$
  correspond to the density matrices $\rho_1,\rho_2$, respectively. The Connes spectral distance
  between $\psi_1$ and $\psi_2$ is defined as
  \begin{equation}
    d (\psi_1, \psi_2) \equiv d (\rho_1, \rho_2) = \sup_{a \in B} |
    \mathrm{tr} (\rho_1 a) - \mathrm{tr} (\rho_2 a) | .
  \end{equation}
  
\end{definition}

If $a\in B$, then we also have $-a\in B$ as well. So there is
\begin{equation}
  d (\rho_1, \rho_2) = \sup_{a \in B} [\mathrm{tr} (\rho_1 a) - \mathrm{tr}
  (\rho_2 a)] = \sup_{a \in B} \left[ \mathrm{tr} (\Delta \hspace{0.17em} a)
  \right] = \mathrm{tr} (\Delta \hspace{0.17em} a_{\ast}),
\end{equation}
where $\Delta = \rho_1 - \rho_2$ is the difference of the states, and $a_{\ast}$
is the so-called \textit{optimal element}. As noted in Ref.~\cite{Iochum}, the supremum in the above Connes spectral distance function can be attained among Hermitian elements; thus it suffices to consider Hermitian optimal elements $a_{\ast}$.
Since $[D,\pi(a+
c\mathbb{I})] = [D, \pi (a)]$,
one may always restrict attention to traceless elements of $B$. Define the set
\begin{equation}
	\mathcal{V} := \{x \in A : x = x^{\dagger}, \mathrm{Tr} x = 0\}, 
\end{equation}
and we have
\begin{equation}
d (\rho_1, \rho_2) = \sup_{a \in \mathcal{V}, L_D (a) \leqslant 1} \left[
   \mathrm{tr} (\Delta \hspace{0.17em} a) \right] .
\end{equation}

In quantum mechanics,
most physical properties are invariant under unitary transformations.
In the framework of noncommutative geometry,
Connes spectral distances between states  can also reflect the relations between states and properties of quantum systems.
It is therefore natural to require the corresponding Connes spectral distance to be unitarily invariant;
that is, for any two states $\rho,\sigma$, one may require
\begin{equation}
  d (U \rho U^{\dagger}, U \sigma U^{\dagger}) = d (\rho, \sigma), \qquad U
  \in \mathcal{U} (A) .
\end{equation}
In quantum information theory,
quantum trace distance and Hilbert-Schmidt distance are very useful measures of distance between two quantum states, which are defined, respectively, as
\begin{equation}
  T (\rho_1, \rho_2) = \frac{1}{2} \| \Delta \|_1, \qquad h (\rho_1, \rho_2) =
  \frac{1}{\sqrt{2}} \| \Delta \|_{\mathrm{HS}}, \label{eq:TH}
\end{equation}
where $\| \Delta \|_1 =\mathrm{Tr} \sqrt{\Delta^{\dag}
\Delta}$, $\| \Delta \|_{\mathrm{HS}} = \sqrt{\mathrm{Tr}
(\Delta^{\dag} \Delta)}$.
It is easy to see that, both distances equal $1$ for orthogonal pure states, and
both of these familiar distances are unitarily invariant.

In the present work, we will construct the corresponding spectral triples based on the quantum trace distance and Hilbert-Schmidt distance.
For finite-dimensional density matrices, one can take $A =\mathbb{M}_n (\mathbb{C})$, $n
\ge 2$. 

Consider the following Hilbert space
\begin{equation}
  H = \mathbb{C}^2_{\chi} \otimes \mathbb{C}^n_1 \otimes \mathbb{C}^n_2 \otimes \mathbb{C}^n_3, \label{eq:H}
\end{equation}
where the subscripts distinguish the different tensor factors.
The algebra $A$ acts on $H$ from the left through the representation $\pi$,
\[ \pi (a) = \mathbb{I}_2 \otimes \mathbb{I}_n \otimes a \otimes \mathbb{I}_n, \quad a\in A . \]
Let $| i \rangle$ denote the standard basis vector whose $i$th component is $1$ and all other components are $0$,
i.e., $|i\rangle=(0,\ldots,1,\ldots,0)^T$ $(1 \leqslant
\mathit{i} \leqslant n)$. Define $E_{ij}=| i \rangle\langle j |$, and there is
\[ E_{i j} | k \rangle   = \delta_{j k} | i \rangle  , \quad E_{i j} E_{k l} =
   \delta_{j k} | i \rangle \langle l | . \]

One can define the following flip operators:
\begin{align}
  F_{12} & = \sum_{i, j} E_{ij} \otimes E_{ji} \otimes \mathbb{I}_n, & F_{13} & =
  \sum_{i, j} E_{ij} \otimes \mathbb{I}_n \otimes E_{ji}, \nonumber\\
  F_{23} & = \sum_{i, j} \mathbb{I}_n \otimes E_{ij} \otimes E_{ji} . &  & 
  \label{eq:flips}
\end{align}
One can verify that $F_{ij}$ can interchange the $i$th and $j$th factors among
the three copies of $\mathbb{C}^n$. For example, consider the vector $| i, j, k \rangle =
| i \rangle   \otimes | j \rangle   \otimes | k \rangle   \in \mathbb{C}^n_1 \otimes
\mathbb{C}^n_2 \otimes \mathbb{C}^n_3$, and there is
\begin{equation*}
  F_{12} | i, j, k \rangle = \sum_{i', j'} E_{i' j'} | i \rangle   \otimes
  E_{j' i'} | j \rangle   \otimes \mathbb{I}_n | k \rangle  = \sum_{i', j'} \delta_{j' i} | i' \rangle   \otimes \delta_{i' j} | j'
  \rangle   \otimes | k \rangle  = | j, i, k \rangle .
\end{equation*}
These $F_{ij}$ are all real, self-adjoint unitary matrices. For example,
\[ F_{12}^{\dag} = \left( \sum_{i, j} E_{ij} \otimes E_{ji} \otimes \mathbb{I}_n
   \right)^{\dag} = \sum_{i, j} E_{j i} \otimes E_{i j} \otimes \mathbb{I}_n = F_{12},
\]
and
\[ F_{12} F_{12}^{\dag} = F_{12}^{\dag} F_{12} = F_{12} F_{12}
   =\mathbb{I}_{n^3} . \]

Furthermore, one can define the following chirality operator
\[ \Gamma = \sigma_3 \otimes \mathbb{I}_{n^3}, \]
and there is $\Gamma^2 = \mathbb{I}$, and
\[ \Gamma \pi (a) = \left( \sigma_3 \otimes \mathbb{I}_{n^3} \right) \left( \mathbb{I}_2
   \otimes \mathbb{I}_n \otimes a \otimes \mathbb{I}_n \right) = \pi (a) \Gamma,
  \]
or $[\Gamma, \pi (a)] = 0$. So this spectral triple is even.

Next, one can also define a real structure. First, let us introduce the following operator
\begin{equation}
  Q = \sigma_1 \otimes F_{23} .
\end{equation}
By the properties of the Pauli matrices and $F_{23}$, there is $Q=\bar Q =Q^\dagger$, $Q^2=\mathbb{I}$, and $Q$ is
unitary,
\[ Q Q^{\dag} = Q Q =\mathbb{I}. \]
Define the real structure as follow,
\begin{equation}
  J = QK, \label{eq:JG}
\end{equation}
where $K$ is the complex-conjugation operator. Then
\[ J^2 = QK QK = Q \bar{Q} = \mathbb{I} . \]
For any vectors $\psi, \varphi \in H$, there is
\[ \langle J \psi, J \varphi \rangle = \langle QK \psi, QK \varphi \rangle =
   \langle K \psi, K \varphi \rangle = \langle \varphi, \psi \rangle . \]
So $J$ is antiunitary.

We also have
\begin{equation}
     J \Gamma 
     = (\sigma_1 \otimes F_{23}) K (\sigma_3 \otimes \mathbb{I}_{n^3})= - (\sigma_3 \otimes \mathbb{I}_{n^3}) (\sigma_1 \otimes F_{23}) K= - \Gamma J.
   \end{equation}
Furthermore, one can also define the right action as
\begin{equation}
  b^{\circ}  = J \pi (b)^{\dagger} J^{- 1} .  \label{eq:right}
\end{equation}
For any vector $\psi \in H$, there is
\begin{eqnarray*}
  b^{\circ} \psi & = & J \pi (b)^{\dagger} J^{- 1} \psi
  = J \pi (b)^{\dagger} J \psi = QK \pi (b)^{\dagger} QK \psi\\
  & = & Q \left( \mathbb{I}_2 \otimes \mathbb{I}_n \otimes b^T \otimes \mathbb{I}_n \right) Q \psi = (\mathbb{I}_2 \otimes \mathbb{I}_n \otimes \mathbb{I}_n \otimes b^T) \psi,
\end{eqnarray*}
or
\begin{equation}
	b^{\circ} = \mathbb{I}_2 \otimes \mathbb{I}_n \otimes \mathbb{I}_n \otimes b^T .
\end{equation}
This is an opposite algebra, i.e., $(bc)^{\circ}=c^{\circ}b^{\circ}$.
Because the left and right actions occupy the second and third $\mathbb{C}^n$
factors, respectively, the order-zero condition $[\pi(a),b^{\circ}]=0$ holds automatically.

\subsection{Flip-Type Dirac Operator and Quantum Trace Distance}
Now we will construct the Dirac operators. First,
let us construct a spectral triple based on the quantum trace distance.

Define the Dirac operator as follow,
\begin{equation}
  D_{\mathrm{tr}} = \frac{1}{\ell} \sigma_1 \otimes (F_{12} + F_{13}), \label{eq:Dtr}
\end{equation}
where the parameter $\ell\in\mathbb{R}$.

Since $F_{ij}=F_{ij}^{\dagger}$, the above Dirac operator is self-adjoint, $D_{\mathrm{tr}}^{\dagger}=D_{\mathrm{tr}}$. From
$\sigma_1\sigma_3=-\sigma_3\sigma_1$, it follows that
\begin{equation}
  D_{\mathrm{tr}} \Gamma = - \Gamma D_{\mathrm{tr}}, \qquad \{ D_{\mathrm{tr}}, \Gamma\} = 0.
\end{equation}
Furthermore, since $D_{\mathrm{tr}}$ is a real matrix, and $F_{23}F_{12}F_{23}=F_{13}$, $F_{23}F_{13}F_{23}=
F_{12}$, one can obtain
\[ J D_{\mathrm{Tr}} = (\sigma_1 \otimes F_{23}) K \cdot
   \frac{1}{\ell} \sigma_1 \otimes (F_{12} + F_{13}) = D_{\mathrm{tr}} J. \]
Since $[F_{13},\mathbb{I}_n\otimes a\otimes\mathbb{I}_n]=0$, we also have
\begin{equation}
  [D_{\mathrm{tr}}, \pi (a)] = \ell^{- 1} \sigma_1 \otimes [F_{12}, \mathbb{I}_n \otimes a
  \otimes \mathbb{I}_n] . \label{eq:flipcomm}
\end{equation}
The right-hand side acts as the identity on the third factor, so the first-order condition $[[D_{\mathrm{tr}},\pi
(a)], b^{\circ}] = 0$ also holds automatically. 

So there is $(\varepsilon,\varepsilon',\varepsilon'')=(+1,+
1,-1)$, and the spectral triple $(A,H,D,\Gamma,J)$ has KO-dimension
$6$. 

Next, let us compute the Lipschitz seminorm.
\[ [D_{\mathrm{tr}}, \pi (a)] = \ell^{- 1} \sigma_1 \otimes [F_{12}, \mathbb{I}_n \otimes a
   \otimes \mathbb{I}_n] = \ell^{- 1} \sigma_1 \otimes \left[ (a \otimes \mathbb{I} - \mathbb{I}
   \otimes a) \otimes \mathbb{I} \right] F_{12} . \]
Since $F_{12}$ and $\sigma_1$ are all unitary, there is
\begin{equation*}
  \| [D_{\mathrm{tr}}, \pi (a)] \|_{op} = \ell^{- 1}  \left\| \sigma_1 \otimes \left[
  (a \otimes \mathbb{I} - \mathbb{I} \otimes a) \otimes \mathbb{I} \right] F_{12}
  \right\|_{op} = \ell^{- 1}  \left\| a \otimes \mathbb{I} - \mathbb{I} \otimes a
  \right\|_{op} .
\end{equation*}
For $a=a^{\dagger}$, the eigenvalues of $a\otimes\mathbb{I}-\mathbb{I}\otimes a$ are
$\lambda_i(a)-\lambda_j(a)$, therefore,
\begin{equation}
  \label{eq:Ltr} L_{\mathrm{tr}} (a) = \| [D_{\mathrm{tr}}, \pi (a)] \|_{op} =
  \frac{\lambda_{\max} (a) - \lambda_{\min} (a)}{\ell} =
  \frac{\mathrm{diamspec} (a)}{\ell} .
\end{equation}
Here ``diamspec($x$)'' means the diameter of the spectrum of $x$,
$
  \mathrm{diamspec} (x) := \lambda_{\max} (x) - \lambda_{\min} (x)$ .
So in this spectral triple, the ball condition $L_{\mathrm{tr}} (a) \leqslant 1$ becomes
\[ \mathrm{diamspec} (a) \leqslant \ell . \]

Define the matrix
\[ M = \frac{1}{\ell} (a - \lambda_{\min} (a)\mathbb{I}), \]
and it is easy to see that, the eigenvalues of $M$ are
\[ 0 \leqslant \frac{1}{\ell} (\lambda_i (a) - \lambda_{\min} (a)) \leqslant
   1. \]
So we have $0 \leqslant M \leqslant \mathbb{I}$.

Since $\mathrm{Tr} \Delta=\mathrm{Tr} (\rho -\sigma) = 0$, we have
\[ \mathrm{Tr}(\Delta a) =\mathrm{Tr}
   [\Delta (a - \lambda_{\min} (a)\mathbb{I})] =\ell \mathrm{Tr}
   [\Delta M] . \]
Therefore, the Connes spectral distance between the states $\rho, \sigma$ in
the spectral triple $(A, H, D_{\mathrm{Tr}}, \Gamma, J)$ is
\begin{equation}
  d_{\mathrm{tr}} (\rho, \sigma) = \sup_{L_{\mathrm{tr}} (a) \leqslant 1}
  \mathrm{Tr} (\Delta a)
   = \ell \sup_{0 \leqslant M \leqslant \mathbb{I}} \mathrm{Tr} (\Delta M) .
  \label{eq:trace-result}
\end{equation}

Consider the decomposition:
\[ \Delta = \Delta_+ - \Delta_-, \]
where $\Delta_{\pm } \geqslant 0$, and $\Delta_+ \Delta_- = 0$. From $\mathrm{Tr}
\Delta = 0$, we have
\[ \mathrm{Tr} \Delta_+ = \mathrm{Tr} \Delta_- = \frac{1}{2} \| \Delta \|_1 . \]
For $0 \leqslant M \leqslant \mathbb{I}$, there is
\[ \mathrm{Tr} (\Delta M) = \mathrm{Tr} (\Delta_+ M) - \mathrm{Tr} (\Delta_- M) \leqslant \mathrm{Tr}
   (\Delta_+ M) \leqslant \mathrm{Tr} \Delta_+ , \]
and the equality is attained by the positive spectral projection $M = P_+
(\Delta)$. So one can obtain
\begin{equation}\label{dtr}
	d_{\mathrm{tr}} (\rho, \sigma) = \ell \sup_{0 \leqslant M \leqslant \mathbb{I}} \mathrm{Tr}
   (\Delta M) = \ell \mathrm{Tr} \Delta_+ = \frac{\ell}{2} \| \Delta \|_1 = \ell T
   (\rho, \sigma) .
\end{equation}

When the parameter $\ell=1$ in the Dirac operator $D_{\mathrm{tr}}$ (\ref{eq:Dtr}), then the spectral distance will return to normal quantum trace distance.

\subsection{Projection-Type Dirac Operator and Hilbert-Schmidt Distance}

Next, we will construct another spectral triple based on the Hilbert-Schmidt distance.
First, define
\begin{equation}
  | \Omega \rangle = \frac{1}{\sqrt{n}}  \sum_{j = 1}^n | j \rangle \otimes | j \rangle,
  \qquad P_{\Omega} = | \Omega \rangle \langle \Omega | = \frac{1}{n}  \sum_{i, j} E_{ij} \otimes
  E_{ij},
\end{equation}
On the three tensor factors, define the operators
\begin{equation}
  \label{eq:Ps} P_{12} = P_{\Omega} \otimes \mathbb{I}_n, \qquad P_{13} = F_{23}
  P_{12} F_{23} = \frac{1}{n}  \sum_{i, j} E_{ij} \otimes \mathbb{I}_n \otimes E_{ij},
\end{equation}
It is easy to see that, these operators $P_{ij}$ are real self-adjoint projections.

Define the Dirac operator as
\begin{equation}
  D_{\mathrm{H}} = \frac{\sqrt{2 n}}{\ell} \sigma_1 \otimes (P_{12} + P_{13}) .
  \label{eq:DH}
\end{equation}
Obviously, there is $D_{\mathrm{H}}^{\dagger} = D_{\mathrm{H}}$ and
$\{ D_{\mathrm{H}}, \Gamma\} = 0$.
Using $P_{13} = F_{23} P_{12} F_{23}$ and $P_{12} = F_{23} P_{13} F_{23}$, one can
obtain
$
F_{23} (P_{12} + P_{13}) = (P_{12} + P_{13}) F_{23}
$,
and then
\[ J D_H = (\sigma_1 \otimes F_{23}) K \cdot \frac{\sqrt{2 n}}{\ell} 
   \hspace{0.17em} \sigma_1 \otimes (P_{12} + P_{13}) = D_H J. \]
Since $P_{13}$ commutes with $\mathbb{I}_n \otimes a \otimes \mathbb{I}_n$, we also have
\begin{eqnarray}
  {}[D_{\mathrm{H}}, \pi (a)] & = & \frac{\sqrt{2 n}}{\ell} [\sigma_1 \otimes
  (P_{12} + P_{13}), \pi (a)] \nonumber\\
  & = & \frac{\sqrt{2 n}}{\ell} \sigma_1 \otimes [P_{12}, \mathbb{I}_n \otimes a
  \otimes \mathbb{I}_n] .  \label{eq:flip-comm}
\end{eqnarray}
The right-hand side acts as the identity on the third factor $\mathbb{C}_3^n$, so there is
\[ [[D_{\mathrm{H}}, \pi (a)], b^{\circ}] = 0. \]

Now let us compute the Lipschitz seminorm. Define $A_a=\mathbb{I}_n\otimes a$. Since $a=
a^{\dagger}$, we also have $A_a = A_a^{\dag}$. It is easy to see that, there is
\[ \langle \Omega | A_a | \Omega \rangle = \frac{1}{n} \mathrm{Tr} a := \tau (a) . \]
Take the orthogonal decomposition
\begin{equation}
  A_a | \Omega \rangle = \tau (a) | \Omega \rangle + | v \rangle, \qquad \langle \Omega |v
  \rangle = 0.
\end{equation}
So there is
$ | v \rangle = A_a | \Omega \rangle - \tau (a) | \Omega \rangle$,
and
\[ \|v\|^2 = \langle v | v  \rangle = \langle \Omega |A_a^2 | \Omega \rangle -
   \tau (a)^2 . \]
Then we have
\begin{equation}
  [P_{\Omega}, A_a] = P_{\Omega} A_a - A_a P_{\Omega} = | \Omega \rangle \langle v | -
  | v \rangle \langle \Omega | .
\end{equation}
If $| v \rangle=0$, the above commutator vanishes; if $| v \rangle\ne0$, one can define
the unit vector $| e \rangle = \frac{| v \rangle}{\|v\|}$. In the base $\left\{
| \Omega \rangle, | e \rangle \right\}$, we have the matrix representation
\[ [P_{\Omega}, A_a] = \|v\| \left(\begin{array}{cc}
     0 & 1\\
     - 1 & 0
   \end{array}\right) . \]
Hence
\begin{equation}\label{pa1}
  \| [P_{\Omega}, A_a] \|_{op}^2 = \|v\|^2 = \langle \Omega |A_a^2 | \Omega
  \rangle - \tau (a)^2 = \tau (a^2) - \tau (a)^2 .
\end{equation}
Denote $a_0 = a - \tau (a)  \mathbb{I}_n$, we have $a_0^{\dag} = a_0$,
$\mathrm{Tr} (a_0) = 0$, and
\[ a_0^{\dag} a_0 = a^2 - 2 \tau (a) a + \tau (a)^2 \mathbb{I}. \]
So there is
\begin{equation}\label{pa2}
\| a_0 \|_{\mathrm{HS}}^2 =\mathrm{Tr} (a_0^{\dag} a_0)
   =\mathrm{Tr} (a^2) - 2 \tau (a)
   \mathrm{Tr} (a) + n \tau (a)^2 = n \tau (a^2) - n \tau
   (a)^2 .
\end{equation}
Combining these two results (\ref{pa1}) and (\ref{pa2}), we have
\[ \| [P_{\Omega}, A_a] \|_{op}^2 = \frac{1}{n} \| a_0 \|_{\mathrm{HS}}^2 . \]
From the result (\ref{eq:flip-comm}), we have
\begin{eqnarray*}
  {}[D_{\mathrm{H}}, \pi (a)] 
  & = & \frac{\sqrt{2 n}}{\ell} \sigma_1 \otimes [P_{\Omega} \otimes \mathbb{I}_n,
  A_a \otimes \mathbb{I}_n]\\
  & = & \frac{\sqrt{2 n}}{\ell} \sigma_1 \otimes [P_{\Omega}, A_a] \otimes
  \mathbb{I}_n.
\end{eqnarray*}
So there is
\begin{equation}
  L_{\mathrm{H}} (a) = \| [D_{\mathrm{H}}, \pi (a)] \|_{op} = \frac{\sqrt{2 n}}{\ell}
  \| [P_{\Omega}, A_a] \|_{op} = \frac{\sqrt{2}}{\ell} \| a_0 \|_{\mathrm{HS}} .
  \label{eq:LH}
\end{equation}

By $\mathrm{Tr} \Delta = 0$ and $a = a_0 + \tau (a)  \mathbb{I}_n$, one can obtain
\begin{eqnarray}\label{dhrs}
  d_{\mathrm{H}} (\rho, \sigma) & = & \sup_{L_{\mathrm{H}} (a) \leqslant 1}
  \left| \mathrm{Tr} (\Delta a) \right|= \sup_{\| a_0 \|_{\mathrm{HS}} \leq \ell / \sqrt{2}} | \mathrm{Tr} (\Delta a_0) |\nonumber\\
  & = & \sup_{\| a_0 \|_{\mathrm{HS}} \leq \ell / \sqrt{2}} | \langle \Delta, a_0
  \rangle_{\mathrm{HS}} | .
\end{eqnarray}
Using the Cauchy-Schwarz inequality $| \langle \Delta, a_0
\rangle_{\mathrm{HS}} | \leqslant \| \Delta \|_{\mathrm{HS}} \cdot
\| a_0 \|_{\mathrm{HS}}$, we have
\[ d_{\mathrm{H}} (\rho, \sigma) \leqslant \sup_{\| a_0 \|_{\mathrm{HS}} \leq \ell /
   \sqrt{2}} \| \Delta \|_{\mathrm{HS}} \cdot \| a_0 \|_{\mathrm{HS}} = \frac{\ell}{\sqrt{2}}
   \| \Delta \|_{\mathrm{HS}} . \]
For example, when $\Delta \ne 0$, one can choose the optimal element
\[ a_0' = \frac{\ell}{\sqrt{2}} \frac{\Delta}{\| \Delta \|_{\mathrm{HS}}}, \]
and there is
\[ L_{\mathrm{H}} (a_0') = \frac{\sqrt{2}}{\ell} \| a_0' \|_{\mathrm{HS}} = 1. \]
Therefore, the Connes spectral distance is
\begin{equation}
  d_{\mathrm{H}} (\rho, \sigma) = \mathrm{Tr} (\Delta a_0') = \frac{\ell}{\sqrt{2}}
  \| \Delta \|_{\mathrm{HS}} = \ell h (\rho, \sigma) . \label{eq:HSdistance}
\end{equation}
Obviously, when $\Delta=0$, there is $d_{\mathrm{H}}(\rho,\sigma)=0$.

When the parameter $\ell=1$ in the Dirac operator $D_{\mathrm{tr}}$ (\ref{eq:DH}), then the spectral distance will return to normal Hilbert-Schmidt distance.

\subsection{Direct-Sum Construction and Generalized Spectral Distance}

Based on the two spectral triples above,
we will construct a further spectral triple with a generalized spectral distance.
Define the new Hilbert space
$\mathcal{H}=H\oplus H$, and construct the following operators:
\begin{equation}
  D_t = D_{\mathrm{tr}} \oplus t D_{\mathrm{H}}, \quad \pi_t = \pi \oplus \pi, \quad \Gamma_t =
  \Gamma \oplus \Gamma, \quad J_t = J \oplus J, \quad t > 0. \label{eq:family}
\end{equation}
It is easy to see that, this new spectral triple $(A,\mathcal{H},D_t,\Gamma_t,J_t)$ is still real and even with KO-dimension 6.

Since the operator norm of a block-diagonal matrix is the maximum of the norms of its blocks, for self-adjoint
$a$, one can obtain
\begin{equation}
  L_t (a) := \| [D_t, \pi_t (a)] \|_{op} = \frac{1}{\ell} \max \{ \mathrm{diamspec} (a),
  \sqrt{2}  \hspace{0.17em} t \| a_0 \|_{\mathrm{HS}} \} . \label{eq:Lt}
\end{equation}
Writing $a_0=a-\tau(a)\mathbb{I}_n$ as $\ell x$, we have
\[ \frac{1}{\ell} \mathrm{diamspec} (a) \leqslant 1 \Longleftrightarrow \mathrm{diamspec} (x)
   \leqslant 1. \]
Denote
\begin{equation}
  r_t = \frac{1}{\sqrt{2} t}, \qquad \mathcal{C}_t = \{x \in \mathcal{V}: \mathrm{diamspec} (x)
  \leqslant 1, \| x \|_{\mathrm{HS}} \leqslant r_t \} .
\end{equation}
Then
\begin{eqnarray*}
  d_t (\rho, \sigma) & = & \sup_{L_t (a) \leqslant 1} \mathrm{Tr} (\Delta a)= \ell \sup_{L_t (a) \leqslant 1} \mathrm{Tr} (\Delta \cdot \frac{1}{\ell}
  \left[ a - \tau (a) \mathbb{I}_n \right])\\
  & = & \ell \sup_{x \in \mathcal{C}_t}  \mathrm{Tr} (\Delta x)\\
  & = & \ell N_t (\Delta),
\end{eqnarray*}
where
\begin{equation}
  N_t (\Delta) \equiv \delta_t (\rho, \sigma) = \frac{d_t (\rho, \sigma)}{\ell} = \sup_{x \in \mathcal{C}_t}  \mathrm{Tr}
  (\Delta x) . \label{eq:Nt}
\end{equation}

From the construction (\ref{eq:family}), one can see that, when the parameters $\ell=1$ and $t=0$, then the spectral distance $d_t (\rho, \sigma)$ will return to normal quantum trace distance between the states $\rho, \sigma$; When the parameters $\ell=t\to\infty$, then the spectral distance $d_t (\rho, \sigma)$ will return to the Hilbert-Schmidt distance.

Since the function $N_t(\Delta)=\delta_t (\rho, \sigma)$ differs from the spectral distance $d_t(\rho,
\sigma)$ defined above only by a constant factor,
one can only study the newly defined spectral-distance function $N_t(\Delta)$ or $\delta_t (\rho, \sigma)$.

Because $\mathcal{C}_t$ is a compact, convex, centrally symmetric set containing a neighborhood of the origin, and
$N_t$ is its support function, $N_t$ is also a norm on $\mathcal{V}$.
So the maximum above can be attained. Thus $\delta_t$
is a finite metric satisfying positivity, symmetry, and the triangle inequality.

It is easy to see that, the set $\mathcal{C}_t$ is invariant under arbitrary unitary conjugations,
\[ x \in \mathcal{C}_t \Longleftrightarrow Ux U^{\dagger} \in \mathcal{C}_t, \]
Therefore, for any $U\in \mathcal{U}(A)$, we have
\begin{eqnarray}
  N_t  (U \Delta U^{\dagger}) & = & \sup_{x \in \mathcal{C}_t}  \mathrm{Tr} (U \Delta
  U^{\dagger} x) 
  = \sup_{x \in \mathcal{C}_t}  \mathrm{Tr} (\Delta U^{\dagger} x U) \nonumber\\
  & = & \sup_{Ux' U^{\dagger} \in \mathcal{C}_t}  \mathrm{Tr} (\Delta x') 
  = \sup_{x' \in \mathcal{C}_t}  \mathrm{Tr} (\Delta x') \nonumber\\
  & = & N_t (\Delta),  \label{eq:UI}
\end{eqnarray}
Thus, in the spectral triple $(A,\mathcal{H},D_t)$,
the Connes spectral distance is also invariant under unitary transformations.

\section{Parameterized Spectral Distances and Computational Methods}\label{sec3}

In this section, we will try to calculate the Connes spectral distances explicitly. To obtain exact results for the spectral distance,
we first need to prove several relevant results.

\begin{lemma}
  \label{lemma1}For every $\Delta\in\mathcal{V}$, we have
  \begin{equation}\label{xyd}
    N_t (\Delta) = \min_{\substack{X, Y \in \mathcal{V}\\ X + Y = \Delta}}
    \left\{ \frac{1}{2} \| X \|_1 + r_t \| Y \|_{\mathrm{HS}} \right\} .
  \end{equation}
 \end{lemma}
 Here $X,Y$
are just traceless self-adjoint matrices, and need not be positive or density matrices.
\begin{proof}
The support function of a convex set $K$ is defined as
\[ h_K (Z) = \underset{x \in K}{\sup } \mathrm{Tr} (Z x)
. \]	
 Let the two convex sets in the traceless Hermitian subspace be $C = \{x :
  \mathrm{diam} \mathrm{spec} (x) \leq 1\}$ and $B = \{x : \|x\|_{\mathrm{HS}} \leq r_t \}$.
So from the results (\ref{eq:trace-result}) and (\ref{dtr}), there is
\begin{equation}
	h_C (X) = \underset{\mathrm{diamspec} (x) \leq 1}{\sup
	} \mathrm{Tr} (X x) = \frac{1}{2} \parallel X
	\parallel_1.
\end{equation}
Similar to the results (\ref{dhrs}) and (\ref{eq:HSdistance}), we also have
\begin{equation}
	h_B (Y) = \underset{\parallel x
		\parallel_{\mathrm{HS}} \leq r_t}{\sup }
	\mathrm{Tr} (Y x)
	= r_t \parallel Y \parallel_{\mathrm{HS}}
	.
\end{equation}
So for any $\Delta = X + Y$ and $x \in C \cap B$, there is
  \begin{eqnarray*}
    \mathrm{Tr} (\Delta x) & = & \mathrm{Tr} [(X + Y) x]
    = \mathrm{Tr} (X x)
    +\mathrm{Tr} (Y x) \nonumber\\
    & \leq & \frac{1}{2} \| X \|_1 + r_t \| Y \|_{\mathrm{HS}} .
  \end{eqnarray*}
  
Let us introduce the indicator function of the set $K$,
  \[ \iota_K (x) = \left\{ \begin{array}{ll}
       0, & x \in K,\\
       + \infty, & x \not\in K,
     \end{array} \right. \]
and we have
$\iota_{C \cap B} = \iota_C + \iota_B $.
The Fenchel conjugate of indicator function is
$$   \iota_K^{\ast} (Z)  = \underset{x \in K}{\sup } \{
       \mathrm{Tr} (Z x) - \iota_K (x) \} = \underset{x \in K}{\sup } \mathrm{Tr} (Z x) = h_K (Z) .$$
Therefore, we have
\[ h_{C \cap B} = \iota_{C \cap B}^{\mathord{*}} = (\iota_C +
     \iota_B)^{\ast} . \]
Using the formula of Fenchel conjugate
$ (f + g)^{\ast} = f^{\ast} \Box g^{\ast}$, where the infimal convolution is defined as
\[ (f \Box g) (\Delta) = \underset{\Delta = X + Y}{\inf }
     [f (X) + g (Y)] , \]
one can obtain
\[       h_{C \cap B} (\Delta)  = (\iota_C + \iota_B)^{\ast} (\Delta) = \iota_C^{\ast} \Box \iota_B^{\ast} (\Delta) = h_C \Box h_B (\Delta),\]
or
  \begin{eqnarray}
    h_{C \cap B} (\Delta) & = & \underset{\Delta = X + Y}{\inf } [h_C (X) +
    h_B (Y)]\nonumber\\
    & = & \underset{\Delta = X + Y}{\inf } \left[ \frac{1}{2} \parallel X
    \parallel_1 + r_t \parallel Y \parallel_{\mathrm{HS}}
    \right] .
  \end{eqnarray}
  This proves the result.
\end{proof}

In the result above, choosing $(X,Y)=(\Delta,0)$ and $(0,
\Delta)$ gives only
\begin{equation}
  N_t (\Delta) \le \min \left\{ T (\rho, \sigma), \frac{h (\rho, \sigma)}{t}
  \right\} . \label{eq:minbound}
\end{equation}
In general, however, this inequality cannot be replaced by an equality, because an optimal decomposition may involve nonzero
$X$ and $Y$ simultaneously.

In general, in the calculations of spectral distances,
finding an optimal element are very difficult.
But if we can only consider diagonal matrices, the problem will become much simpler.

\begin{theorem}
  \label{thm1}In the spectral triple $(A, \mathcal{H},
  D_t)$, for any two states $\rho, \sigma \in A$, and $\Delta = \rho -
  \sigma$, there is
  \begin{equation}\label{eq:pinching-reduction}
    N_t (\Delta) =  \sup_{x \in \mathcal{C}_t}
    \mathrm{Tr} (\Delta x) = \sup_{\substack{x \in \mathcal{C}_t\\ [x, \Delta] = 0}}
    \mathrm{Tr} (\Delta x) .
  \end{equation}
  So the corresponding optimal element can always be chosen to commute with
  $\Delta$.
  
  Furthermore, if
  \[ \Delta = V\mathrm{diag}(\lambda_1, \ldots, \lambda_n)
     V^{\dagger}, \qquad V \in \mathcal{U} (A), \]
  then there exists an optimal element of the form
  \[ x_{\ast} = V\mathrm{diag} (x_1, \ldots, x_n)
     V^{\dagger} . \]
\end{theorem}

\begin{proof}
Since $\Delta$ is Hermitian, we may write $\Delta = V \mathrm{diag}
(\lambda_1, \ldots, \lambda_n) V^{\dagger} = V \Delta' V^{\dagger}$, where
$\Delta' = \mathrm{diag} (\lambda_1, \ldots, \lambda_n)$. Since the set
$\mathcal{C}_t$ is also invariant under unitary transformations, there is
\begin{eqnarray}\label{diagd}
	N_t (\Delta) & = & \sup_{x \in \mathcal{C}_t} \mathrm{Tr} (\Delta x) =
	\sup_{V^{\dagger} x V \in \mathcal{C}_t} \mathrm{Tr} (V \Delta' V^{\dagger}
	x) \nonumber\\
	& = & \sup_{V^{\dagger} x V \in \mathcal{C}_t} \mathrm{Tr} (\Delta'
	V^{\dagger} x V) = \sup_{x' \in \mathcal{C}_t} \mathrm{Tr} (\Delta' x') . 
\end{eqnarray}
Therefore, for simplicity, one can simply consider $\Delta$ as a traceless
diagonal real matrix in the spectral distance function, and set $\Delta
\rightarrow \Delta' = \mathrm{diag} (\lambda_1, \ldots, \lambda_n)$.

Let $Q_j = E_{jj} = |i \rangle \langle i|$, and there is $Q_j^{\dagger} =
Q_j,$
\[ Q_i Q_j = \delta_{i j} Q_j, \qquad \sum_{j = 1}^n Q_j =\mathbb{I}, \qquad
\Delta = \sum_{j = 1}^n \lambda_j Q_j . \]
Define the following map $\mathcal{P} (x)$ for any matrix $x$, we have
\begin{equation}
	\label{px} \mathcal{P} (x) = \sum_{j = 1}^n Q_j xQ_j =
	\sum_{j = 1}^n E_{jj} xE_{jj} = \mathrm{diag} (x_{11}, \ldots, x_{nn}) .
\end{equation}
Obviously, there is
\begin{equation}
	\label{trpx} \mathrm{Tr} \mathcal{P} (x) = \mathrm{Tr} x.
\end{equation}
For the two diagonal matrices $\Delta, \mathcal{P} (x)$, there is
\begin{equation}
	\mathrm{Tr} (\Delta \mathcal{P}(x)) = \sum_j \lambda_j x_{j j} = \sum_j
	\lambda_j \mathrm{Tr} (xQ_j) = \mathrm{Tr} \left( x \sum_j \lambda_j Q_j
	\right) = \mathrm{Tr} (x \Delta) .
\end{equation}
For any matrix $x$, we have
\begin{equation}
	\label{pxhs} \|\mathcal{P}(x)\|_{\mathrm{HS}}^2 = \sum_i |x_{ii} |^2 \leqslant
	\sum_{i, j} |x_{i j} |^2 = \|x\|_{\mathrm{HS}}^2 .
\end{equation}
Therefore, if $\|x\|_{\mathrm{HS}} \le r_t$, we also have
$\|\mathcal{P}(x)\|_{\mathrm{HS}} \le r_t$.

Let $m = \lambda_{\min} (x)$ and $M = \lambda_{\max} (x)$. Since $x =
x^{\dagger}$, we have the operator inequality
\[ m\mathbb{I} \leqslant x \leqslant M\mathbb{I}. \]
It is easy to see that, the map $\mathcal{P}$ is positive and unital. Applying
$\mathcal{P}$ to the above inequality, we obtain
\[ m\mathbb{I}=\mathcal{P} (m\mathbb{I}) \leqslant \mathcal{P} (x) \leqslant
\mathcal{P} (M\mathbb{I}) = M\mathbb{I}. \]
Therefore,
\[ m \leqslant \lambda_{\min} (\mathcal{P}(x)) \leqslant \lambda_{\max}
(\mathcal{P}(x)) \leqslant M. \]
Hence,
\begin{equation}
	\label{diapx} \mathrm{diamspec} (\mathcal{P}(x)) = \lambda_{\max}
	(\mathcal{P}(x)) - \lambda_{\min} (\mathcal{P}(x)) \leqslant M - m =
	\mathrm{diamspec} (x) .
\end{equation}
Thus, if $\mathrm{diamspec} (x) \le 1$, then we also have $\mathrm{diamspec}
(\mathcal{P}(x)) \leqslant 1$.

Define the set
\[ \mathcal{C}_t' = \left\{ x \in A : \mathcal{P}(x)
=\mathcal{P}(x)^{\dagger}, \mathrm{Tr} \mathcal{P}(x) = 0, 
\mathrm{diamspec} (\mathcal{P}(x)) \leq 1, 
\|\mathcal{P}(x)\|_{\mathrm{HS}} \le r_t \right\}, \]
\[ \mathcal{C}_t'' = \{x = \mathrm{diag} (x_1, \ldots, x_n) \in A : x =
x^{\dagger}, \mathrm{Tr} x = 0, \mathrm{diamspec} (x) \leqslant 1,
\|x\|_{\mathrm{HS}} \leqslant r_t \} . \]
Obviously, we have $\mathcal{C}_t'' \subseteq \mathcal{C}_t \subseteq
\mathcal{C}_t'$. So there is
\begin{eqnarray}
	N_t (\Delta) & = & \sup_{x \in \mathcal{C}_t} \mathrm{Tr} (\Delta x)
	\leqslant \sup_{x \in \mathcal{C}_t'} \mathrm{Tr} (\Delta x) \nonumber\\
	& = & \sup_{x \in \mathcal{C}_t'} \mathrm{Tr} (\Delta \mathcal{P}(x)) =
	\sup_{x \in \mathcal{C}_t''} \mathrm{Tr} (\Delta x) \nonumber\\
	& \leqslant & \sup_{x \in \mathcal{C}_t} \mathrm{Tr} (\Delta x) . 
\end{eqnarray}
So there must be
\[ N_t (\Delta) = \sup_{x \in \mathcal{C}_t''} \mathrm{Tr} (\Delta x) . \]
This means that, in the spectral distance function, if $\Delta$ is diagonal,
one only need to consider the diagonal optimal elements in the set
$\mathcal{C}_t$.

So in general, if we have $\Delta = V \mathrm{diag} (\lambda_1, \ldots,
\lambda_n) V^{\dagger}$, then there exists an corresponding optimal element of
the form $x_{\ast} = V \mathrm{diag} (x_1, \ldots, x_n) V^{\dagger}$.
Obviously, there is $[x_{\ast}, \Delta] = 0$.
\end{proof}

\begin{theorem}
  \label{prop:pinching}Let $\Delta=V\mathrm{diag}(\lambda_1,\ldots,\lambda_n)
  V^{\dagger}$. Then
  \begin{equation}
    N_t (\Delta) = \sup_{x \in W} \sum_{i = 1}^n \lambda_i x_i
    \label{eq:socp}
  \end{equation}
  where the set
  \[ W := \left\{ x = (x_1, \ldots, x_n) \in \mathbb{R}^n : \sum_{i = 1}^n
     x_i = 0, c \le x_i \le c + 1, \|x\|_2 \le r_t , c\in\mathbb{R}. \right\}, \]
  and $\|x\|_2:=\sqrt{x_1^2+\cdots+x_n^2}$.
\end{theorem}

\begin{proof}
  By Theorem (\ref{thm1}), an optimal element $x$ can always be chosen to commute with $\Delta$,
  and thus can be expressed as
  \begin{equation*}
    x = V \mathrm{diag} (x_1, \ldots, x_n) V^{\dagger}, \qquad x_i \in \mathbb{R}.
  \end{equation*}
  Therefore, $\mathrm{Tr} x=0$ yields
  \begin{equation*}
    \sum_{i = 1}^n x_i = 0.
  \end{equation*}
  The constraint $\|x\|_{\mathrm{HS}}\le r_t$ gives
  \begin{equation*}
    \sqrt{x_1^2 + \cdots + x_n^2} \le r_t .
  \end{equation*}
Since $\mathrm{spec}(x)=\{x_1,\ldots,x_n\}$, the condition $\mathrm{diam}\mathrm{spec}(x)\le
  1$ implies that
$ \max x_i - \min x_i \le 1$.
So there exists $c\in\mathbb{R}$ such that
  \begin{equation*}
    c \le x_i \le c + 1, \qquad i = 1, \ldots, n.
  \end{equation*}
Furthermore, there is
\begin{equation*}
    \mathrm{Tr} (\Delta x)  = \mathrm{Tr} \left[ V \mathrm{diag} (\lambda_1, \ldots, \lambda_n) V^{\dagger} V \mathrm{diag} (x_1, \ldots, x_n) V^{\dagger} \right]  = \sum_{i = 1}^n \lambda_i x_i . 
  \end{equation*}
Hence,
  \[ N_t (\Delta) = \sup_{\substack{x = x^{\dagger}, \mathrm{Tr} x = 0 \\\mathrm{diam}  \mathrm{spec}
     (x) \le 1 \\\|x\|_{\mathrm{HS}} \le r_t}}  \mathrm{Tr} (\Delta x) = \sup_{x \in
     W} \sum_{i = 1}^n \lambda_i x_i . \]
  
\end{proof}

This result means that, in this case, the Connes spectral distance only depends on the spectrum of $\Delta$.
In general, one may always diagonalize $\Delta$ first,
and then restrict the corresponding optimal element $x_{\ast}$ to diagonal, real, traceless matrices.

Lemma~\ref{lemma1} can also be reduced to the following form.

\begin{corollary}
 Suppose $\Delta=V\mathrm{diag}(\lambda_1,\ldots,\lambda_n)V^{\dagger}$, and denote $\lambda=(\lambda_1,\ldots,\lambda_n)^T$, then
  \begin{equation}
    N_t (\Delta) = \min_{\substack{z \in \mathbb{R}^n\\\sum_i z_i = 0}}
    \left\{ \frac{1}{2}  \sum_{i = 1}^n |z_i | + r_t \| \lambda - z\|_2
    \right\} . \label{eq:socpdual}
  \end{equation}
\end{corollary}

\begin{proof}
Similar to the result (\ref{diagd}), one can simply assume that $\Delta$ being diagonal, and set $\Delta=\mathrm{diag}(\lambda_1,\ldots,\lambda_n)$.
 Consider any decomposition in the calculation (\ref{xyd}),
  \[ \Delta = X + Y. \]
 Similar to the construction (\ref{px}), one can define the map $\mathcal{P}$ and apply to both sides of the above equality, 
  \begin{equation}
    \mathcal{P} (\Delta) = \Delta =\mathcal{P} (X)
    +\mathcal{P} (Y) .
  \end{equation}
 Denote the diagonal matrices as
 $ X' =\mathcal{P} (X)$, $Y' =\mathcal{P} (Y) $. 
  Hence,
  \[ X' + Y' = \Delta, \quad [X', \Delta] = [Y', \Delta] = 0. \]
  Moreover,
  \[ \mathrm{Tr} X' = \mathrm{Tr} X = 0, \qquad \mathrm{Tr} Y' = \mathrm{Tr} Y = 0. \]
  Thus, $(X',Y')$ remains a feasible decomposition in (\ref{xyd}).
 
Since $X$ is Hermitian, one can consider the following spectral decomposition
\[ X = \sum_j u_j |e_j \rangle \langle e_j |, \]
and there is
\[ \|X\|_1 = \sum_j | u_j | . \]
\[ x_{i i} = \langle i| X |i \rangle = \langle i| \sum_j u_j |e_j \rangle
\langle e_j | |i \rangle = \sum_j u_j \langle i| e_j \rangle \langle e_j |i
\rangle \]
So we have
\begin{eqnarray}
		\|X' \|_1 & = & \sum_i | x_{i i} | = \sum_i \left| \sum_j u_j \langle i|e_j \rangle \langle e_j |i \rangle
		\right| \nonumber\\
		& \leqslant & \sum_i \sum_j | u_j | \langle i| e_j \rangle \langle e_j |i
		\rangle = \sum_j \sum_i | u_j | \langle e_j |i \rangle \langle i| e_j \rangle
		\nonumber\\
		& = & \sum_j | u_j | = \|X\|_1 . 
\end{eqnarray}
We also have
  \begin{equation}
    \|Y' \|_{\mathrm{HS}}^2 = \sum_i |y_{ii} |^2 \leqslant
    \sum_{i, j} |y_{i j} |^2=
    \|Y\|_{\mathrm{HS}}^2 .
  \end{equation}
  Therefore,
  \begin{equation}
    \frac{1}{2}  \|X' \|_1 + r_t \|Y' \|_{\mathrm{HS}} \le \frac{1}{2} \|X\|_1
    + r_t \|Y\|_{\mathrm{HS}} .
  \end{equation}

  Thus, an optimal solution of the distance function  (\ref{xyd}) can always be chosen so that $X, Y$ being diagonal, and
$\Delta=\mathrm{diag}(\lambda_1,\ldots,\lambda_n)=X+Y$.
Denote the optimal matrices as
  \begin{equation*}
    X = \mathrm{diag} (z_1, \ldots, z_n),
\quad    Y =\Delta-X= \mathrm{diag} (\lambda_1 - z_1, \ldots, \lambda_n - z_n) .
  \end{equation*}

  The condition $\mathrm{Tr} X=0$ implies
  \begin{equation*}
    \sum_{i = 1}^n z_i = 0.
  \end{equation*}
  On the other hand, since $\sum_i\lambda_i=0$ and $\sum_i z_i=0$,
  we automatically have $\mathrm{Tr} Y=0$. Therefore, we have
  \[ \|X\|_1 = \sum_{i = 1}^n |z_i |, 
  \qquad \|Y\|_{\mathrm{HS}} 
  = \left( \sum_{i = 1}^n
     (\lambda_i - z_i)^2 \right)^{1 / 2} = \| \lambda - z\|_2 . \]
  Hence, the distance function (\ref{xyd}) can be expressed as
  \begin{equation}
    N_t (\Delta) = \min_{\substack{z \in \mathbb{R}^n\\\sum_i z_i = 0}}
    \left\{ \frac{1}{2}  \sum_{i = 1}^n |z_i | + r_t \| \lambda - z\|_2
    \right\} .
  \end{equation}
\end{proof}

\section{Global Parameter Regimes}\label{sec4}
Obviously, in the spectral triple $(A,\mathcal{H},D_t)$, the spectral distance $N_t(\Delta):=
\delta_t(\rho,\sigma)$ depends on the parameter $t$.
In this section, we will study the formulas in different parameter regimes.

\begin{lemma}
  For any Hermitian matrix $a\in M_n(\mathbb{C})$, denote
  \[ a_0 = a - \frac{\mathrm{Tr} a}{n} \mathbb{I}_n , \]
then there is
  \begin{equation}
    \label{eq:diam-lower} \mathrm{diamspec} (a) \le \sqrt{2} \hspace{0.17em} \|a_0
    \|_{\mathrm{HS}} .
  \end{equation}
\end{lemma}

\begin{proof}
  Let the eigenvalues of $a$ be $\lambda_1,\lambda_2,\ldots,\lambda_n\in
  \mathbb{R}$. Denote their mean by
  \begin{equation*}
    \bar{\lambda} = \frac{1}{n}  \sum_{i = 1}^n \lambda_i = \frac{1}{n}  \mathrm{Tr}
    a.
  \end{equation*}
  The eigenvalues of $a_0=a-\bar\lambda \mathbb{I}_n$ are $\lambda_i-
  \bar\lambda$, $i=1,\ldots,n$. Therefore,
  \begin{equation*}
    \label{eq:HS-centered} \|a_0 \|_{\mathrm{HS}}^2 = \mathrm{Tr} (a_0^2) = \sum_{i =
    1}^n (\lambda_i - \bar{\lambda})^2 .
  \end{equation*}
  On the other hand, by the definition of spectral diameter,
\begin{equation*}
    \left( \mathrm{diam}  \mathrm{spec} (a) \right)^2  = (\lambda_{\max} - \lambda_{\min})^2
    = [(\lambda_{\max} - \bar{\lambda}) - (\lambda_{\min} - \bar{\lambda})]^2
    . 
  \end{equation*}
Using the elementary inequality $(u-v)^2\le2(u^2+v^2)$, we obtain
  \begin{eqnarray*}
    \left( \mathrm{diam}  \mathrm{spec} (a) \right)^2 & \leqslant & 2 [(\lambda_{\max} -
    \bar{\lambda})^2 + (\lambda_{\min} - \bar{\lambda})^2]\\
    & \leqslant & 2 \sum_{i = 1}^n (\lambda_i - \bar{\lambda})^2= 2 \|a_0 \|_{\mathrm{HS}}^2 .
  \end{eqnarray*}
Hence, there is 
$
    \mathrm{diam}  \mathrm{spec} (a) \le \sqrt{2} \hspace{0.17em} \|a_0 \|_{\mathrm{HS}} $.

\end{proof}

\begin{lemma}
  Let $a=a^{\dagger}\in \mathbb{M}_n(\mathbb{C})$ and $ a_0 = a - \frac{\mathrm{Tr} a}{n} \mathbb{I}_n $,
 then we have
  \begin{equation}
    \label{eq:diam-upper} \|a_0 \|_{\mathrm{HS}} \le C_n  \hspace{0.17em}
    \mathrm{diamspec} (a), \qquad C_n = \sqrt{\frac{\lfloor n^2 / 4 \rfloor}{n}} .
  \end{equation}
  Equivalently,
  \[ C_n = \left\{\begin{array}{ll}
       \dfrac{\sqrt{n}}{2}, & n\text{ is even},\\
       \frac{1}{2}\sqrt{n-\frac{1}{n}}, & n\text{ is odd}.
     \end{array}\right. \]
\end{lemma}

\begin{proof}
  Let the eigenvalues of $a$ be $\lambda_1,\ldots,\lambda_n\in\mathbb{R}$. Denote
  \[ m = \lambda_{\min} (a), \qquad M = \lambda_{\max} (a), \qquad d = M - m =
     \mathrm{diam}  \mathrm{spec} (a) . \]
  If $d=0$, then $a=c\mathbb{I}_n$ for some $c\in\mathbb{R}$ and consequently
  $a_0=0$. In this case the equality holds.
  
  Now let us assume $d>0$. Define
  \begin{equation}
    u_i = \frac{\lambda_i - m}{d}, \qquad i = 1, \ldots, n.
  \end{equation}
  Since $m\le\lambda_i\le M$, we have $0\le u_i\le1$. Thus $\lambda_i
  =m+du_i$. Denote
  \[ \bar{\lambda} = \frac{1}{n}  \sum_{i = 1}^n \lambda_i, \qquad \bar{u} =
     \frac{1}{n}  \sum_{i = 1}^n u_i . \]
  Then we have
  \[ \bar{\lambda} = \frac{1}{n}  \sum_i (m + d u_i) = m + d \bar{u} . \]
  Therefore,
  \begin{equation*}
    \lambda_i - \bar{\lambda} = d (u_i - \bar{u}) .
  \end{equation*}
  Since the eigenvalues of $a_0$ are just $\lambda_i-\bar\lambda$, there is
\begin{equation*}
    \|a_0 \|_{\mathrm{HS}}^2 = \sum_{i = 1}^n (\lambda_i - \bar{\lambda})^2
    = d^2  \sum_{i = 1}^n (u_i - \bar{u})^2 . 
\end{equation*}
Since
\begin{equation*}
    \sum_{i = 1}^n (u_i - \bar{u})^2 = \sum_i u_i^2 - 2 \bar{u}  \sum_i u_i
    + n \bar{u}^2  = \sum_i u_i^2 - \frac{1}{n} \left( \sum_i u_i \right)^2 ,
\end{equation*}
therefore, we have
  \begin{equation}
    \label{eq:variance-cube} \|a_0 \|_{\mathrm{HS}}^2 = d^2  \left[ \sum_{i =
    1}^n u_i^2 - \frac{1}{n} \left( \sum_{i = 1}^n u_i \right)^2 \right] .
  \end{equation}
  Define the function
  \begin{equation}
    F (u_1, \ldots, u_n) = \sum_{i = 1}^n u_i^2 - \frac{1}{n} \left( \sum_{i =
    1}^n u_i \right)^2 = \mathbf{u}^T  \left( \mathbb{I}- \frac{1}{n}  \textbf{v v}^T
    \right) \mathbf{u},
  \end{equation}
  where $\mathbf{u}=(u_1,\ldots,u_n)^T$ and $\mathbf{v}=(1,\ldots,1)^T$. The matrix $\mathbb{I}-
  \frac{1}{n}\mathbf{v}\mathbf{v}^T$ is the orthogonal projector onto the subspace orthogonal to $\mathbf{v}$,
  and is therefore positive semidefinite. Hence $F$
  is convex. Since $[0,1]^n$ is a compact convex polytope,
  a convex function attains its maximum over this polytope at some vertex.
  The vertices of $[0,1]^n$ are exactly those with $u_i\in\{0,1\}$, $i=1,\ldots,n$.
  
  Suppose a vertex has exactly $k$ components equal to $1$ and the remaining $n-k$
  components equal to $0$. Then we have
  \[ \sum_{i = 1}^n u_i = k, \qquad \sum_{i = 1}^n u_i^2 = k, \]
 and
  \begin{equation*}
    F (u) = k - \frac{k^2}{n}  = \frac{k (n - k)}{n} = \frac{1}{n} \left[ - \left( k - \frac{n}{2}
    \right)^2 + \frac{n^2}{4} \right] . 
  \end{equation*}
Therefore, $F(u)$ is maximized when $k$ is as close as possible to $n/2$.
  
  If $n$ is even, then $k=n/2$, and
  \[ \max_k k (n - k) = \frac{n^2}{4} . \]
  If $n$ is odd, then $k=(n-1)/2$ $\text{or}$ $k=
  (n+1)/2$, and
  \[ \max_k k (n - k) = \frac{n^2 - 1}{4} . \]
Finally, we have
  \begin{equation}
    \label{eq:Fmax} F (u) \le \frac{\lfloor n^2 / 4 \rfloor}{n} .
  \end{equation}
From \ref{eq:variance-cube} and \ref{eq:Fmax}, one can obtain
\begin{equation}
    \|a_0 \|_{\mathrm{HS}}^2 \leqslant d^2  \frac{\lfloor n^2 / 4 \rfloor}{n} = \frac{\lfloor n^2 / 4 \rfloor}{n} \left( \mathrm{diam}  \mathrm{spec} (a) \right)^2 .
  \end{equation}
namely,
  \begin{equation}
    \|a_0 \|_{\mathrm{HS}} \le \sqrt{\frac{\lfloor n^2 / 4 \rfloor}{n}} 
    \hspace{0.17em} \mathrm{diamspec} (a) .
  \end{equation}
  
\end{proof}

\subsection{Three Distance Regimes}

Denote
\begin{equation*}
  \label{eq:global-threshold} t_{\mathrm{c}} (n) = \frac{1}{\sqrt{2} 
  \hspace{0.17em} C_n} = \sqrt{\frac{n}{2 \lfloor n^2 / 4 \rfloor}},
\end{equation*}
and consider the following three regimes for the parameter $t$.

\subsubsection{Case I: $0<t\le t_{\mathrm{c}}(n)$}

In this case,
\begin{equation}
  C_n \le \frac{1}{\sqrt{2}  \hspace{0.17em} t} = r_t .
\end{equation}
Thus, whenever $\mathrm{diam}\mathrm{spec}(x)\le1$, we have
\begin{equation}
  \|x\|_{\mathrm{HS}} \le C_n  \mathrm{diam}  \mathrm{spec} (x) \le C_n \le r_t .
\end{equation}
In other words, in this parameter regime,
the spectral-diameter constraint automatically implies the HS-ball constraint.
So we have
\begin{equation}
  \mathcal{C}_t = \left\{ x = x^{\dagger} : \mathrm{Tr} x = 0, \mathrm{diam}  \mathrm{spec} (x) \le 1
  \right\} ,
\end{equation}
and
\begin{equation}
	\label{ntd}
  N_t (\Delta) = \sup_{\substack{x = x^{\dagger}, \mathrm{Tr} x = 0 \\\mathrm{diam}  \mathrm{spec} (x)
  \le 1}}  \mathrm{Tr} (\Delta x) .
\end{equation}
It is easy to see that, this is equivalent to choosing the following Dirac operator:
\begin{equation}
 D= \ell D_{\mathrm{tr}} = \sigma_1 \otimes (F_{12} + F_{13}) ,
\end{equation}
and
\[ \| [D, \pi (a)] \|_{op} =\| [\ell D_{\mathrm{tr}}, \pi (a)] \|_{op} =\mathrm{diamspec} (a)
   \leqslant 1. \]
Thus, from the results (\ref{eq:trace-result}) and (\ref{dtr}), the spectral distance (\ref{ntd}) is
\[ N_t (\Delta)=\delta_t (\rho, \sigma) = \frac{1}{\ell} d_{\mathrm{tr}} (\rho, \sigma) =
   \sup_{0 \leqslant M \leqslant \mathbb{I}} \mathrm{Tr} (\Delta M) = \frac{1}{2}
   \| \Delta \|_1 = T (\rho, \sigma) . \]
So in the small-$t$ regime, the parameterized Connes spectral
distance reduces exactly to the standard quantum trace distance.

\subsubsection{Case II: $t\ge1$}

If the HS-ball constraint holds,
\begin{equation*}
  \|x\|_{\mathrm{HS}} \le r_t = \frac{1}{\sqrt{2}  \hspace{0.17em} t},
\end{equation*}
Eq.~(\ref{eq:diam-lower}) implies
\begin{equation*}
  \mathrm{diamspec} (x) \leqslant \sqrt{2} \hspace{0.17em} \|x\|_{\mathrm{HS}} \leqslant \sqrt{2}  \hspace{0.17em} r_t = \frac{1}{t} \leqslant 1. 
\end{equation*}
So in this case, we have
\begin{equation}\label{ntd1}
  N_t (\Delta) = \sup_{\substack{x = x^{\dagger}, \mathrm{Tr} x = 0
  \\\|x\|_{\mathrm{HS}} \le r_t}}  \mathrm{Tr} (\Delta x) .
\end{equation}
Similarly, this is equivalent to choosing the following Dirac operator:
\begin{equation}
 D= \frac{\ell}{r_t \sqrt{2}} D_{\mathrm{H}} = \frac{\sqrt{n}}{r_t} \sigma_1 \otimes
  (P_{12} + P_{13}) , 
\end{equation}
and
\[ \| [D, \pi (x)] \|_{op} = \| [\frac{\ell}{r_t \sqrt{2}} D_{\mathrm{H}}, \pi (x)] \|_{op} = \frac{1}{r_t}
   \| x \|_{\mathrm{HS}} \leqslant 1. \]
Thus, from the results (\ref{dhrs}) and (\ref{eq:HSdistance}), the spectral distance (\ref{ntd1}) is
\[ N_t (\Delta)=\delta_t (\rho, \sigma) = \frac{r_t \sqrt{2}}{\ell} d_{\mathrm{H}} (\rho,
   \sigma) = r_t \| \Delta \|_{\mathrm{HS}} = r_t \sqrt{2} h (\rho, \sigma) = \frac{1}{t}
   h (\rho, \sigma) . \]
So in the large-$t$ regime, the spectral distance is determined entirely by the Hilbert--Schmidt ball.

\subsubsection{Case III: $t_{\mathrm{c}}(n)<t<1$}

In this intermediate regime, both constraints may restrict the set:
\begin{equation}
  \mathcal{C}_t = \left\{ x : \mathrm{diam}  \mathrm{spec} (x) \le 1 \right\} \cap \{ x :
  \|x\|_{\mathrm{HS}} \le r_t \} .
\end{equation}
In general, one must therefore compute
\begin{equation}
  N_t (\Delta) = \sup_{\substack{x = x^{\dagger}, \mathrm{Tr} x = 0 \\\mathrm{diam}  \mathrm{spec} (x)
  \le 1 \\\|x\|_{\mathrm{HS}} \le r_t}}  \mathrm{Tr} (\Delta x),
\end{equation}
and the spectral distance is determined jointly by the intersection of the two convex sets.

In summary, the Connes spectral distances for the different values of $t$ are
\begin{equation}
  \label{eq:global-regimes} \delta_t (\rho, \sigma) = \left\{\begin{array}{ll}
    T (\rho, \sigma), & 0 < t \leq t_{\mathrm{c}} (n),\\[.5em]
    \text{generic intersection-constrained distance}, & t_{\mathrm{c}} (n) < t
    < 1,\\[.5em]
    \dfrac{1}{t} h (\rho, \sigma), & t \geq 1.
  \end{array}\right.
\end{equation}
For some low-dimensional examples, the values of $t_{\mathrm{c}}$ are
\begin{equation}
  t_{\mathrm{c}} (2) = 1, \qquad t_{\mathrm{c}} (3) = \frac{\sqrt{3}}{2},
  \qquad t_{\mathrm{c}} (4) = \frac{1}{\sqrt{2}} .
\end{equation}

Since
\begin{equation}
  r_t = \frac{1}{\sqrt{2} t}, \qquad \mathcal{C}_t = \{x \in \mathcal{V}: \mathrm{diamspec} (x)
  \leqslant 1, \| x \|_{\mathrm{HS}} \leqslant r_t \} .
\end{equation}
as $t$ increases, $r_t$ decreases, and hence the set $\mathcal{C}_t$ becomes smaller.
Therefore, for any fixed states $\rho,\sigma$, the corresponding spectral distance
\begin{equation}
  N_t (\Delta) := \delta_t (\rho, \sigma) = \sup_{x \in \mathcal{C}_t}  \mathrm{Tr} (\Delta x)
\end{equation}
is monotonically nonincreasing in $t$.

\section{Exact Thresholds Depending on the Difference Spectrum}\label{sec5}
Since the properties of spectral distances may also depend on the structure of the states or $\Delta$, we will analysis the relation between the exact thresholds of the parameter $t$ and $\Delta$.

\begin{theorem}
  \label{thm:pairthresholds}Let $k,l$ be the numbers of positive and negative eigenvalues of $\Delta=\rho-\sigma$,
  respectively, and let $m=k+l$;
  zero eigenvalues are not included in $m$. Define
  \begin{equation}
    \label{eq:pairthresholds} t_- (\Delta) = \sqrt{\frac{m}{2 kl}}, \qquad t_+
    (\Delta) = \frac{\lambda_{\max} (\Delta) - \lambda_{\min}
    (\Delta)}{\sqrt{2} \hspace{0.17em} \| \Delta \|_{\mathrm{HS}}} .
  \end{equation}
  Then
  \begin{equation}
    t_{\mathrm{c}} (n) \le t_- (\Delta) \le t_+ (\Delta) \le 1.
  \end{equation}
  Moreover,
  \begin{equation}
    \label{eq:pairregimes} \delta_t (\rho, \sigma) = T (\rho, \sigma)
    \Longleftrightarrow t \le t_- (\Delta),
  \end{equation}
  and
  \begin{equation}
    \delta_t (\rho, \sigma) = \frac{h (\rho, \sigma)}{t} \Longleftrightarrow t
    \ge t_+ (\Delta) .
  \end{equation}
  If
  \[ t_- (\Delta) < t < t_+ (\Delta), \]
  then both the spectral-diameter and the Hilbert--Schmidt
  ball constraints on the optimal observable are saturated, that is,
  \begin{equation}
  	\mathrm{diam}  \mathrm{spec} (x_{\ast}) = 1, \qquad \|x_{\ast} \|_{\mathrm{HS}} = r_t, 
  	\end{equation}
  and
  \begin{equation}
    \delta_t (\rho, \sigma) < \min \left\{ T (\rho, \sigma), \frac{h (\rho,
    \sigma)}{t} \right\} .
  \end{equation}
\end{theorem}

\begin{proof}
  Let the positive eigenvalues of $\Delta$ be $\alpha_1,\ldots,\alpha_k>0$,
  and the negative eigenvalues be $-\beta_1,\ldots,-\beta_l$, where $\beta_j>0$.
  The remaining $n-m$ eigenvalues are zero.
Since $\mathrm{Tr}\Delta=0$, we have
  \begin{equation}
    \sum_{i = 1}^k \alpha_i = \sum_{j = 1}^l \beta_j := S.
  \end{equation}
  Since
$ \| \Delta \|_1 = \sum_i \alpha_i + \sum_j \beta_j = 2 S$,
  it follows that
  \begin{equation}
    T (\rho, \sigma) = \frac{1}{2} \| \Delta \|_1 = S.
  \end{equation}

Let us analysis the first critical parameter $t_-$.

First consider
\begin{equation}
    \sup_{\substack{x = x^{\dagger}, \mathrm{Tr} x = 0 \\\mathrm{diam}  \mathrm{spec} (x) \le 1}}  \mathrm{Tr}
    (\Delta x) .
  \end{equation}
  We have already shown that its optimal value is
$ T(\rho, \sigma) = S$.
  In an eigenbasis of $\Delta$, denote $x=\mathrm{diag}(x_1,\ldots,x_n)$. Let $M=
  \max_i x_i$ and $q=\min_i x_i$. The spectral-diameter constraint gives $M-q\le1$.
  Hence,
\begin{equation}
    \mathrm{Tr} (\Delta x) = \sum_{i = 1}^k \alpha_i x_i - \sum_{j = 1}^l \beta_j
    x_{k + j} \leqslant M \sum_{i = 1}^k \alpha_i - q \sum_{j = 1}^l \beta_j = S (M - q) \leqslant S. 
  \end{equation}
  
  Thus, to attain the trace-distance optimum $\mathrm{Tr}(\Delta x)=S=T(\rho, \sigma)$,
  the inequalities above must be equalities. Every optimal element attaining the trace distance
  $T$ must therefore take constant values $M$ and $q$ on the positive and negative
  spectral subspaces of $\Delta$, respectively, with $M-q
  = 1$.

  For any optimal element attaining $T$, its Hilbert--Schmidt norm satisfies at least
  \begin{equation}
    \|x\|_{\mathrm{HS}}^2 \ge kM^2 + lq^2,
  \end{equation}
  because the squared contributions from the zero-eigenvalue directions are nonnegative.
 Using $q=M-1$ and $m=k+l$, one can obtain
\begin{equation}
    kM^2 + lq^2 
    = m \left( M - \frac{l}{m} \right)^2 + \frac{kl}{m} .
  \end{equation}
Therefore,
  \begin{equation}
    \|x\|_{\mathrm{HS}}^2 \ge \frac{kl}{m} .
  \end{equation}
  The minimum is attained when
  \[ M = \frac{l}{m}, \qquad q = - \frac{k}{m} . \]
For example, the optimal element can be chosen as
  \begin{equation}
    x_T = \mathrm{diag} \left( \underbrace{\frac{l}{m}, \ldots, \frac{l}{m}}_k,
    \underbrace{- \frac{k}{m}, \ldots, - \frac{k}{m}}_l, \underbrace{0,
    \ldots, 0}_{n - m} \right) .
  \end{equation}
It is easy to see that, there is
$ \mathrm{Tr} x_T = 0$, and $ \mathrm{diam}  \mathrm{spec} (x_T) = 1$.
  Its HS norm is
\begin{equation}
    \|x_T \|_{\mathrm{HS}}^2  = k \frac{l^2}{m^2} + l \frac{k^2}{m^2} = \frac{kl (k + l)}{m^2} = \frac{kl}{m} . 
  \end{equation}
Hence, in this case, there is
  \begin{equation}
    \min \|x\|_{\mathrm{HS}} =
    \sqrt{\frac{kl}{m}} . \label{eq:minHS-traceopt}
  \end{equation}

  The full spectral distance also requires
  \[ \|x\|_{\mathrm{HS}} \le r_t = \frac{1}{\sqrt{2}  \hspace{0.17em} t} . \]
  Therefore, there exists an optimal element satisfying the HS constraint and attaining the trace distance $T$
  if and only if
  \begin{equation}
    \sqrt{\frac{kl}{m}} \le \frac{1}{\sqrt{2}  \hspace{0.17em} t} ,
  \end{equation}
 or
  \begin{equation}
    t \le \sqrt{\frac{m}{2 kl}} = t_- (\Delta) .
  \end{equation}
So there is
  \begin{equation}
    \delta_t (\rho, \sigma) = T (\rho, \sigma) \Longleftrightarrow t \le t_-
    (\Delta) .
  \end{equation}
  
Next, let us calculate the second critical parameter $t_+$.

  Now consider only
  \begin{equation}
    \sup_{\substack{x = x^{\dagger}, \mathrm{Tr} x = 0 \\\|x\|_{\mathrm{HS}} \le r_t}}
    \mathrm{Tr} (\Delta x) .
  \end{equation}
  By the Hilbert-Schmidt Cauchy-Schwarz inequality, there is
 \begin{equation}
    \mathrm{Tr} (\Delta x) \le \| \Delta \|_{\mathrm{HS}} \cdot\|x\|_{\mathrm{HS}} \le r_t \| \Delta \|_{\mathrm{HS}} = \frac{h (\rho, \sigma)}{t}. 
  \end{equation}
If $\Delta\ne0$, the equalities are attained by
  \begin{equation}
    x_H = r_t  \frac{\Delta}{\| \Delta \|_{\mathrm{HS}}}.
    \label{eq:HSoptimizer}
  \end{equation}
For the full problem, it must also satisfy
$ \mathrm{diamspec} (x_H) \le 1$.
So there is
  \begin{equation}
    \mathrm{diam}  \mathrm{spec} (x_H)  = r_t  \frac{\mathrm{diam}  \mathrm{spec} (\Delta)}{\| \Delta
    \|_{\mathrm{HS}}} = \frac{\lambda_{\max} (\Delta) - \lambda_{\min} (\Delta)}{\sqrt{2} 
    \hspace{0.17em} t \hspace{0.17em} \| \Delta \|_{\mathrm{HS}}}  = \frac{t_+ (\Delta)}{t} \le 1,
  \end{equation}
where
  \[ t_+ (\Delta) = \frac{\lambda_{\max} (\Delta) - \lambda_{\min}
     (\Delta)}{\sqrt{2} \hspace{0.17em} \| \Delta \|_{\mathrm{HS}}} . \]
Hence $\mathrm{diam}\mathrm{spec}(x_H)\le1$ if and only if
$ t \ge t_+ (\Delta) $.
  Therefore, we have
  \begin{equation}
    \delta_t (\rho, \sigma) = \frac{h (\rho, \sigma)}{t} \Longleftrightarrow t
    \ge t_+ (\Delta) .
  \end{equation}
  
From the result (\ref{eq:diam-lower}),
for every Hermitian matrix $a$,
$\mathrm{diamspec} (a) \le \sqrt{2} \hspace{0.17em} \|a_0 \|_{\mathrm{HS}} $.
  Since $\mathrm{Tr}\Delta=0$, we have $\Delta_0=\Delta$. Therefore,
  \[ \lambda_{\max} (\Delta) - \lambda_{\min} (\Delta) \le \sqrt{2}
     \hspace{0.17em} \| \Delta \|_{\mathrm{HS}} . \]
  Hence,
  \begin{equation}
    t_+ (\Delta) = \frac{\mathrm{diam}  \mathrm{spec} (\Delta)}{\sqrt{2} \hspace{0.17em} \|
    \Delta \|_{\mathrm{HS}}} \le 1.
  \end{equation}
  
Since $k+l=m$, the integer product $kl$ is maximized when $k$ and $l$
  are as close as possible, so there is
  \begin{equation}
    kl \le \left\lfloor \frac{m^2}{4} \right\rfloor .
  \end{equation}
Hence, we have
\begin{equation}
    t_-^2  = \frac{m}{2 kl} \ge \frac{m}{2 \lfloor m^2 / 4 \rfloor} = t_{\mathrm{c}} (m)^2 . 
  \end{equation}
On the other hand, $m\le n$, and therefore
$ t_{\mathrm{c}} (m) \ge t_{\mathrm{c}} (n) $. Thus, we have
  \begin{equation}
    t_{\mathrm{c}} (n) \le t_- (\Delta) .
  \end{equation}

Finally, let us prove that $t_-(\Delta)\le t_+(\Delta)$.
  
Denote $A=\lambda_{\max}(\Delta)>0$ and $B=-\lambda_{\min}(\Delta)>0$.
  Then $\mathrm{diamspec}(\Delta)=A+B$, and $0<\alpha_i\le A$ and $0<\beta_j\le B$. So we have
\begin{equation}
    \| \Delta \|_{\mathrm{HS}}^2  = \sum_i \alpha_i^2 + \sum_j \beta_j^2 \leqslant \sum_{i = 1}^k A \alpha_i + \sum_{j = 1}^l B \beta_j
    \le (A + B) S.  \label{eq:HS-AS}
  \end{equation}
On the other hand,
  \[ S = \sum_i \alpha_i \le kA, \quad S = \sum_j \beta_j \le lB, \]
or
  \begin{equation}
    S \le \min \{kA, lB\} \le \frac{kl}{k + l}  (A + B) .
  \end{equation}
Combining with \eqref{eq:HS-AS}, we have
  \begin{equation}
    \| \Delta \|_{\mathrm{HS}}^2 \le \frac{kl}{m}  (A + B)^2 
    =\frac{kl}{m}  (\mathrm{diamspec}(\Delta))^2.
  \end{equation}
Thus,
\begin{equation}
    t_+^2 (\Delta) = \frac{(\mathrm{diamspec}(\Delta))^2}{2 \| \Delta \|_{\mathrm{HS}}^2} \ge \frac{m}{2 kl} = t_-^2(\Delta) . 
  \end{equation}

  Combining the above results, one can obtain
  \begin{equation}
    t_{\mathrm{c}} (n) \le t_- (\Delta) \le t_+ (\Delta) \le 1.
  \end{equation}
  
Now, let us study the intermediate regime between $t_- (\Delta)$ and $t_+ (\Delta)$.

Suppose that there is some parameter $t$ with $t_-(\Delta)<t<t_+(\Delta)$.
Since $t>t_-(\Delta)$,
  the necessary-and-sufficient condition established above implies $\delta_t\ne T$.
  On the other hand, we always have $\delta_t\le T$, so there should be
$ \delta_t < T$.
  Similarly, since $t<t_+(\Delta)$, we have $\delta_t\ne h/t$.
But $\delta_t\le h/t$, so we also have $\delta_t < \frac{h}{t}$.
Hence, there should be
  \begin{equation}
    \delta_t < \min \left\{ T, \frac{h}{t} \right\} .
  \end{equation}
  
Next, let us study the constraints in the intermediate regime.
  
  Let $x_{\ast}$ be an optimal solution of the full problem:
  \[ \delta_t = \mathrm{Tr} (\Delta x_{\ast}) . \]
If $\|x_{\ast}\|_{\mathrm{HS}}<r_t$. Since $\delta_t<T$,
  $x_{\ast}$ does not optimize the spectral-diameter-only problem. Let $x_T$
  be a spectral-diameter-feasible point attaining the trace distance $T$.
Consider
  \[ x_{\varepsilon} = (1 - \varepsilon) x_{\ast} + \varepsilon x_T, \qquad 0
     < \varepsilon \ll 1. \]
  Since the spectral-diameter unit ball is convex, $\mathrm{diam}\mathrm{spec}(x_{\varepsilon})\le1$.
  Because $x_{\ast}$ lies in the interior of the HS ball, continuity of the norm ensures that
  $\|x_{\varepsilon}\|_{\mathrm{HS}}<r_t$ still holds for sufficiently small
  $\varepsilon>0$. Thus $x_{\varepsilon}$ remains feasible.
  
  However, there is
\begin{equation}
    \mathrm{Tr} (\Delta x_{\varepsilon})  = (1 - \varepsilon) \delta_t + \varepsilon
    T  > \delta_t, 
  \end{equation}
which contradicting the optimality of $x_{\ast}$.
  
  Therefore, there should be
  \begin{equation}
    \|x_{\ast} \|_{\mathrm{HS}} = r_t .
  \end{equation}

Similarly, if $\mathrm{diamspec}(x_{\ast})<1$. Since $\delta_t<h/t$,
  $x_{\ast}$ is not the unique optimizer of the HS-ball-only problem, namely $x_H=r_t
  \frac{\Delta}{\|\Delta\|_{\mathrm{HS}}}$. Consider
  \[ x_{\varepsilon} = (1 - \varepsilon) x_{\ast} + \varepsilon x_H . \]
  Since the HS ball is convex, $\|x_{\varepsilon}\|_{\mathrm{HS}}\le r_t$.
  Furthermore, because the spectral diameter is continuous, for sufficiently small $\varepsilon>0$,
  $\mathrm{diam}\mathrm{spec}(x_{\varepsilon})<1$. Thus $x_{\varepsilon}$
  remains feasible for the full problem.
  
  But we also have
\begin{equation}
    \mathrm{Tr} (\Delta x_{\varepsilon})  = (1 - \varepsilon) \delta_t + \varepsilon
    \frac{h}{t}  > \delta_t, 
  \end{equation}
which also contradicting the optimality.
   Therefore, there should be
  \begin{equation}
    \mathrm{diam}  \mathrm{spec} (x_{\ast}) = 1.
  \end{equation}

So in the strict intermediate regime $t_-<t<t_+$,
  every optimal element must simultaneously satisfy
  \begin{equation}
    \mathrm{diam}  \mathrm{spec} (x_{\ast}) = 1, \qquad \|x_{\ast} \|_{\mathrm{HS}} = r_t .
  \end{equation}
  that is, both the spectral-diameter and Hilbert--Schmidt-ball boundaries are active.
  
  This completes the proof.
\end{proof}

In some cases, if
$ t_- (\Delta) = t_+ (\Delta)$,
then the strict intermediate interval is empty.
For example, when all positive eigenvalues of $\Delta$ are equal,
and all negative eigenvalues are also equal, namely,
\[ \mathrm{spec} (\Delta) = \{ \underbrace{A, \ldots, A}_k, \underbrace{- B, \ldots, -
   B}_l, 0, \ldots, 0\}, \]
with the tracelessness condition $kA=lB$.
In this case, all the inequalities in the above proof become equalities, and there is
$ t_-(\Delta) = t_+(\Delta) $.

\section{Spectral Distance Between Arbitrary Pairs of Pure States}\label{sec6}

In general, the calculations of Connes spectral distances are still very cumbersome. Here we will focus on some simple examples.

\begin{proposition}
  \label{cor:ranktwo}Suppose the nonzero eigenvalues of $\Delta$ are $s$ and $-s$,
  where $s\ge0$. Then
  \begin{equation}
    \label{eq:ranktwo-distance} N_t (\Delta) = \alpha_t s,
  \end{equation}
  where $\alpha_t=\min\{1,t^{-1}\}$. In particular, for any two pure states
  $|\psi\rangle,|\phi\rangle$, let $P_{\psi}=| \psi \rangle\langle \psi |$,
  $P_{\phi}=| \phi \rangle\langle \phi |$. Then the spectral distance is
  \begin{equation}
    \label{eq:purepair} \delta_t (P_{\psi}, P_{\phi}) = \alpha_t  \sqrt{1 - |
    \langle \psi | \phi \rangle |^2} .
  \end{equation}
\end{proposition}

\begin{proof}
  If $s=0$, the statement is immediate. So one can assume $ s > 0$.
  
In this case, we have $k=l=1$ and $m=k+l=2$. By Theorem~\ref{thm:pairthresholds},
  the first critical parameter is
  \[ t_- (\Delta) = \sqrt{\frac{m}{2 kl}} = 1. \]
  On the other hand, the second critical parameter is
\begin{equation}
    t_+ (\Delta)  = \frac{\lambda_{\max} (\Delta) - \lambda_{\min}
    (\Delta)}{\sqrt{2} \hspace{0.17em} \| \Delta \|_{\mathrm{HS}}} = \frac{2 s}{\sqrt{2}  (\sqrt{2}  \hspace{0.17em} s)} = 1 = t_- (\Delta) . 
  \end{equation}
So in this case, there is no nonempty intermediate regime $ t_- < t < t_+ $.

  Next, let us compute the trace distance and the normalized Hilbert--Schmidt distance. We have
\begin{equation}
    T  = \frac{1}{2} \| \Delta \|_1 = s, 
\qquad
    h = \frac{1}{\sqrt{2}} \| \Delta \|_{\mathrm{HS}}  = s. 
  \end{equation}
Hence, there is
$ T = h = s$.

  By Theorem~\ref{thm:pairthresholds},
when $0<t\le1$, $N_t(\Delta)=T=s$; when $t\ge1$, $N_t
  (\Delta)=h/t=s/t$. Thus, we have
  \begin{equation}
    N_t (\Delta)  = \alpha_t s= \left\{\begin{array}{ll}
      s, & 0 < t \le 1,\\[.5em]
      \dfrac{s}{t}, & t \ge 1,
    \end{array}\right.
  \end{equation}
where $\alpha_t=\min\left\{1,\frac1t\right\}$.

Next, let us consider the pure states.
Let $\Delta=P_{\psi}-P_{\phi}=| \psi \rangle
  \langle \psi |-| \phi \rangle\langle \phi |$. As $\Delta$ is Hermitian
  and traceless, its two possibly nonzero eigenvalues are of the form $\pm q$.
So we have
\begin{equation}
	\mathrm{Tr} \Delta^2 = q^2 + (- q)^2 = 2 q^2 .
\end{equation}  
There is
\begin{equation}
    \Delta^2  = (P_{\psi} - P_{\phi})^2 = P_{\psi} + P_{\phi} - P_{\psi} P_{\phi} - P_{\phi} P_{\psi} .
    \end{equation}
Taking the trace gives  
  \begin{equation*}
    \mathrm{Tr} \Delta^2  = \mathrm{Tr} P_{\psi} + \mathrm{Tr} P_{\phi} - \mathrm{Tr} (P_{\psi} P_{\phi}) -
    \mathrm{Tr} (P_{\phi} P_{\psi}) = 2 - 2 \mathrm{Tr} (P_{\psi} P_{\phi}) = 2 - 2 | \langle \psi | \phi \rangle |^2 .   \end{equation*}
Combining the above results, one can obtain
  \begin{equation}
    q = \sqrt{1 - | \langle \psi | \phi \rangle |^2} .
  \end{equation}
Thus, for the pure states, there is
  \begin{equation}
    \delta_t (P_{\psi}, P_{\phi}) = \alpha_t  \sqrt{1 - | \langle \psi | \phi
    \rangle |^2} .
  \end{equation}
\end{proof}

  For two pure states, $T(P_{\psi},P_{\phi})=\sqrt{1-|\langle\psi|
  \phi\rangle|^2}$ is precisely the standard pure-state quantum trace distance.
  
  Hence, when $0<t\le1$, there is
  \[ \delta_t (P_{\psi}, P_{\phi}) = T (P_{\psi}, P_{\phi}) . \]
When $t>1$, the spectral distance becomes
  \[ \delta_t (P_{\psi}, P_{\phi}) = \frac{1}{t} T (P_{\psi}, P_{\phi}) . \]

  If $\psi$ and $\phi$ are orthogonal, $\langle\psi|\phi\rangle=0$, so
  \[ \delta_t (P_{\psi}, P_{\phi}) = \alpha_t
  = \left\{\begin{array}{ll}
       1, & 0 < t \le 1,\\
       1 / t, & t \ge 1.
     \end{array}\right. \]

When $\psi$ and $\phi$ differ only by a global phase, $|\langle\psi|
  \phi\rangle|=1$, so there is $P_{\psi}=P_{\phi}$, and
  \[ \delta_t (P_{\psi}, P_{\phi}) = 0. \]
  Thus, from the standpoint of spectral distance, $\psi$ and
  $\phi$ are indistinguishable in this case.

\section{Discriminating Quantum States: Differences from the Ordinary Trace Distance}\label{sec7}

Similar to the quantum trace distance, the Connes spectral distance can also
characterize the relationship between two quantum states
and can be regarded as a quantum distance measure. In the spectral triples constructed above,
the Connes spectral distance generalizes the normal trace and HS distances.
For different values of the parameter $t$,
the Connes spectral distance can differ numerically from the trace distance. In some situations,
it can distinguish state pairs that have the same trace distance.

In this section, we will give a concrete example showing that two pairs of states can have exactly the same quantum trace distance
$T$ and normalized Hilbert--Schmidt distance
$h$ while, in certain parameter regimes, still having different parameterized spectral distances
$\delta_t$.

For example, let us consider the following two traceless Hermitian difference matrices:
\begin{align}
  \Delta_{\mathrm{A}} & = \mathrm{diag} \left( \frac{9}{20}, \frac{1}{20}, -
  \frac{1}{4}, - \frac{1}{4} \right),  \label{eq:DeltaA}\\
  \Delta_{\mathrm{B}} & = \mathrm{diag} \left( \frac{1}{4} + \frac{1}{5 \sqrt{2}},
  \frac{1}{4} - \frac{1}{5 \sqrt{2}}, - \frac{1}{4} + \frac{1}{5 \sqrt{2}}, -
  \frac{1}{4} - \frac{1}{5 \sqrt{2}} \right) ,  \label{eq:sameTHspectra}
\end{align}
and the corresponding pairs of quantum states can be chosen as
\begin{equation}
  \rho_j = \frac{I_4}{4} + \frac{\Delta_j}{2}, \qquad \sigma_j = \frac{I_4}{4}
  - \frac{\Delta_j}{2}, \qquad j = A, B.
\end{equation}
It is easy to check that, the two state pairs have the same trace and Hilbert--Schmidt
distances:
\begin{equation}
  T_{\mathrm{A}} = T_{\mathrm{B}} = \frac{1}{2}, \qquad h_{\mathrm{A}} =
  h_{\mathrm{B}} = \frac{\sqrt{66}}{20} . \label{eq:sameTH}
\end{equation}
Thus, from the viewpoints of trace distance and HS distance,
these two state pairs may be considered to have similar properties.

Next, let us compute their Connes spectral distances.

For the first state pair, the difference matrix have two positive and two negative eigenvalues, so
$ k = l = 2, \, m = 4$.
By Theorem~\ref{thm:pairthresholds}, there is
\begin{equation}
  t_- (\Delta_{\mathrm{A}})  = \sqrt{\frac{m}{2 kl}} = \sqrt{\frac{4}{2 \cdot
  2 \cdot 2}} = \frac{1}{\sqrt{2}} \simeq 0.707107. 
\end{equation}
\begin{equation}
  t_+ (\Delta_{\mathrm{A}})  = \frac{\mathrm{diam}  \mathrm{spec}
  (\Delta_{\mathrm{A}})}{\sqrt{2} \hspace{0.17em} \| \Delta_{\mathrm{A}}
  \|_{\mathrm{HS}}} = \frac{7 / 10}{\sqrt{2}  \hspace{0.17em} \sqrt{33} / 10} = \frac{7}{\sqrt{66}} \simeq 0.861640. 
\end{equation}

Now take $t=0.9>t_+(\Delta_{\mathrm{A}})$.
Thus the first state pair has already entered its Hilbert--Schmidt regime.
By Theorem~\ref{thm:pairthresholds},
we have
\begin{equation}
  N_{0.9} (\Delta_{\mathrm{A}})  = \frac{h_{\mathrm{A}}}{0.9} = \frac{\sqrt{66} / 20}{9 / 10} = \frac{\sqrt{66}}{18} \simeq 0.45133547. 
\end{equation}

For the second state pair, we also have $k=l=2$, $m=4$. Therefore,
\begin{equation}
  t_- (\Delta_{\mathrm{B}}) = \frac{1}{\sqrt{2}} .
\end{equation}
However, $t_+$ differs from that of the first pair.
\[ \mathrm{diam}  \mathrm{spec} (\Delta_{\mathrm{B}}) = \frac{5 + 2 \sqrt{2}}{10} ,\quad \| \Delta_{\mathrm{B}} \|_{\mathrm{HS}} = \frac{\sqrt{33}}{10}, \]
and we obtain
\begin{equation}
  t_+ (\Delta_{\mathrm{B}})  = \frac{\mathrm{diam}  \mathrm{spec}
  (\Delta_{\mathrm{B}})}{\sqrt{2} \hspace{0.17em} \| \Delta_{\mathrm{B}}
  \|_{\mathrm{HS}}} = \frac{(5 + 2 \sqrt{2}) / 10}{\sqrt{66} / 10} = \frac{5 + 2 \sqrt{2}}{\sqrt{66}} \simeq 0.9636. 
\end{equation}
Therefore, at $t=0.9$, the second state pair remains in its strict intermediate regime,
\[ t_- (\Delta_{\mathrm{B}}) < t < t_+ (\Delta_{\mathrm{B}}) . \]
So both the spectral-diameter and HS-ball constraints must be saturated.

Using the reflection symmetry of the eigenvalue problem, an optimal element can be chosen as
\begin{equation}
  x = \mathrm{diag}(u, v, - v, - u), \qquad 0 \le v \le u.
\end{equation}
For such a element, the tracelessness condition holds automatically. The spectral-diameter constraint gives
\begin{equation}
  u \le \frac{1}{2} . \label{eq:B-u-bound}
\end{equation}
On the other hand, the HS-ball constraint is
\[ \|x\|_2 = \sqrt{2 (u^2 + v^2)} \le r_t = \frac{1}{\sqrt{2}  \hspace{0.17em}
   t}, \]
Therefore, we have
\begin{equation}
  u^2 + v^2 \le \frac{1}{4 t^2} . \label{eq:B-circle}
\end{equation}
For the optimal element, we must have
\begin{equation}
  u_{\ast} = \frac{1}{2} , \quad u_{\ast}^2 + v_{\ast}^2 = \frac{1}{4 t^2}.
\end{equation}
Thus, there is
\begin{equation}
  v_{\ast}  = \frac{1}{2}  \sqrt{\frac{1}{t^2} - 1} = \frac{\sqrt{19}}{18} . 
\end{equation}
So the optimal element is
\begin{equation}
  x_{\mathrm{B}}^{\ast} = \mathrm{diag}\left( \frac{1}{2}, \frac{\sqrt{19}}{18}, -
  \frac{\sqrt{19}}{18}, - \frac{1}{2} \right) .
\end{equation}
Substituting $u_{\ast}$ and $v_{\ast}$ into the objective function gives
\begin{equation}
  N_{0.9} (\Delta_{\mathrm{B}}) = \mathrm{Tr} (\Delta_{\mathrm{B}}
  x_{\mathrm{B}}^{\ast}) = \left( \frac{1}{4} + \frac{1}{5 \sqrt{2}} \right) + \left( \frac{1}{4}
  - \frac{1}{5 \sqrt{2}} \right)  \frac{\sqrt{19}}{9} \simeq 0.44400839. 
\end{equation}

In summary, the two state pairs have exactly the same trace and Hilbert--Schmidt distances,
but at the same parameter value $t=0.9$, there is
\begin{equation}
  N_{0.9} (\Delta_{\mathrm{A}}) \simeq 0.45133547, \qquad
  N_{0.9} (\Delta_{\mathrm{B}}) \simeq 0.44400839.  \label{eq:sameTHresult}
\end{equation}
Therefore, in the spectral triple constructed above,
the spectral distances between these two pairs of quantum states differ.

Note that this does not mean the quantum trace distance fails to distinguish distinct quantum states.
The trace distance $T(\rho,\sigma)$ remains a genuine metric on quantum states:
\[ T (\rho, \sigma) = 0 \Longleftrightarrow \rho = \sigma . \]
The point illustrated here is different: two distinct state pairs
$(\rho_{\mathrm{A}}, \sigma_{\mathrm{A}}), (\rho_{\mathrm{B}},
\sigma_{\mathrm{B}})$ can have the same values with $T,h$,
while the parameterized spectral distance $\delta_t$
still distinguishes them geometrically through the finer distribution of the eigenvalues
of their difference matrices.

\section{A Geometric Measure of Entanglement Based on Spectral Distance}\label{sec8}

Since quantum distance measures can be used to characterize the entanglement of quantum states \cite{Vedral}, one can also use Connes spectral distances to analysis quantum entanglement.

The set of separable two-qubit states is defined by
\begin{equation}
  \mathcal{S}= \left\{ \sum_j p_j \rho_j^A \otimes \rho_j^B : p_j \ge 0,
  \sum_j p_j = 1 \right\} .
\end{equation}
In finite dimensions, this is a compact convex set.
By virtue of the method in Ref.~\cite{Vedral}, one can define an type of entanglement measure based on Connes spectral distances.

\begin{definition}
The spectral-distance geometric entanglement measure is defined as
  \begin{equation}
    \mathcal{E}_t (\rho) = \min_{\sigma \in \mathcal{S}} N_t  (\rho - \sigma) .
    \label{eq:Esep}
  \end{equation}
  It represents the shortest spectral distance from the quantum state $\rho$ to the set of separable states.
\end{definition}

\begin{proposition}
  $\mathcal{E}_t$ satisfies the following properties:
  \begin{enumerate}
  	\item[(1).] $\mathcal{E}_t (\rho) \ge 0, \qquad \mathcal{E}_t (\rho) = 0 \Longleftrightarrow \rho
  	\in \mathcal{S};$
    \item[(2).] $\mathcal{E}_t  ((U_A \otimes U_B) \rho (U_A \otimes
    U_B)^{\dagger}) = \mathcal{E}_t (\rho) ;$
    \item[(3).] $\mathcal{E}_t \left( \sum_j p_j \rho_j \right) \le \sum_j p_j
    \mathcal{E}_t (\rho_j) ;$
    \item[(4).] $\left| \mathcal{E}_t (\rho) - \mathcal{E}_t (\eta) \right| \le N_t  (\rho - \eta) . 
    \label{eq:lipschitz}$
  \end{enumerate}
\end{proposition}

\begin{proof}
(1).
  If $\rho\in\mathcal{S}$, then clearly $\mathcal{E}_t(\rho)=0$. Conversely, if
  $ \mathcal{E}_t (\rho) = 0$. then there exists $\sigma_{\rho}^{\ast}\in\mathcal{S}$ such that
  \[ N_t  (\rho - \sigma_{\rho}^{\ast}) = 0. \]
  By positive definiteness of the norm, $\rho-\sigma_{\rho}^{\ast}=0$. Thus,
  $\rho=\sigma_{\rho}^{\ast}\in\mathcal{S}$. Hence,
  \begin{equation}
    \mathcal{E}_t (\rho) = 0 \Longleftrightarrow \rho \in \mathcal{S}.
  \end{equation}

(2).
  The norm $N_t$ is invariant under unitary conjugation, and the set of separable states is also invariant under local unitary transformations. So there is
\begin{eqnarray*}
    \mathcal{E}_t (U \rho U^{\dagger}) && = \min_{\tau \in \mathcal{S}} N_t  (U \rho
    U^{\dagger} - \tau)  = \min_{U \sigma U^{\dagger} \in \mathcal{S}} N_t  (U \rho U^{\dagger}
    - U \sigma U^{\dagger}) \nonumber\\
    && = \min_{\sigma \in \mathcal{S}} N_t  (U (\rho - \sigma) U^{\dagger}) 
     = \min_{\sigma \in \mathcal{S}} N_t  (\rho - \sigma) 
     = \mathcal{E}_t (\rho) . 
  \end{eqnarray*}
Therefore, we have
 \begin{equation}
    \mathcal{E}_t ((U_A \otimes U_B) \rho (U_A \otimes U_B)^{\dagger}) = \mathcal{E}_t (\rho)
    .
  \end{equation}

(3).
  Consider quantum states $\rho_j$ and a probability distribution $p_j\ge0$ with $\sum_jp_j=
  1$. For each $\rho_j$, we can choose a nearest separable state $\sigma_j^{\ast}
  \in\mathcal{S}$ satisfying
  \begin{equation}
    \mathcal{E}_t (\rho_j) = N_t  (\rho_j - \sigma_j^{\ast}) .
  \end{equation}
  Since the set $\mathcal{S}$ of separable states is convex, there is
  \begin{equation}
    \bar{\sigma} = \sum_j p_j \sigma_j^{\ast} \in \mathcal{S}.
  \end{equation}
  Now consider the mixture
  \[ \bar{\rho} = \sum_j p_j \rho_j . \]
  Since $\bar\sigma$ is a valid separable state, it need not be
  the closest separable state to $\bar\rho$, so we have
\begin{eqnarray*}
    \mathcal{E}_t (\bar{\rho}) && = \min_{\sigma \in \mathcal{S}} N_t  (\bar{\rho} -
    \sigma) \le N_t  (\bar{\rho} - \bar{\sigma}) \\
    && = N_t  \left( \sum_j p_j \rho_j - \sum_j p_j \sigma_j^{\ast} \right)  = N_t \left( \sum_j p_j (\rho_j - \sigma_j^{\ast}) \right) . 
  \end{eqnarray*}
Using the triangle inequality and positive homogeneity of the norm, there is
\begin{eqnarray*}
    N_t \left( \sum_j p_j (\rho_j - \sigma_j^{\ast}) \right) && \le \sum_j N_t 
    (p_j (\rho_j - \sigma_j^{\ast})) = \sum_j p_j N_t  (\rho_j - \sigma_j^{\ast}) \\
    && = \sum_j p_j \mathcal{E}_t (\rho_j) . 
  \end{eqnarray*}
Therefore, we have
  \begin{equation}
    \mathcal{E}_t \left( \sum_j p_j \rho_j \right) \le \sum_j p_j \mathcal{E}_t (\rho_j) .
  \end{equation}

(4).
  Choose a nearest separable state $\sigma_{\eta}^{\ast}\in\mathcal{S}$ for $\eta$, so that
  \begin{equation}
    \mathcal{E}_t (\eta) = N_t  (\eta - \sigma_{\eta}^{\ast}) .
  \end{equation}
We also have
  \begin{eqnarray*}
    \mathcal{E}_t (\rho) & \leqslant & N_t  (\rho - \sigma_{\eta}^{\ast})= N_t  [(\rho - \eta) + (\eta - \sigma_{\eta}^{\ast})]\\
    & \leqslant & N_t  (\rho - \eta) + N_t  (\eta - \sigma_{\eta}^{\ast})= N_t  (\rho - \eta) + \mathcal{E}_t (\eta) .
  \end{eqnarray*}
  Therefore,
  \begin{equation}
    \mathcal{E}_t (\rho) - \mathcal{E}_t (\eta) \le N_t  (\rho - \eta) .
    \label{eq:lipschitz-one}
  \end{equation}
Similarly, exchanging $\rho$ and $\eta$ gives 
  \begin{equation}
    \mathcal{E}_t (\eta) - \mathcal{E}_t (\rho) \le N_t  (\eta - \rho) = N_t  (\rho - \eta) .
    \label{eq:lipschitz-two}
  \end{equation}
  Combining \eqref{eq:lipschitz-one} and \eqref{eq:lipschitz-two}, we obtain
  \begin{equation}
    \left| \mathcal{E}_t (\rho) - \mathcal{E}_t (\eta) \right| \le N_t  (\rho - \eta) .
  \end{equation}
\end{proof}

\subsection{Spectral Geometric Entanglement of Bell-Diagonal States}
Now, we will compute the entanglement of quantum states in several concrete examples.
For comparison, we will also consider some other measures such as concurrence and negativity.

Consider a two-qubit system. In the computational basis $\{| 00 \rangle,| 01 \rangle,| 10 \rangle,
| 11 \rangle\}$, define the states
\begin{equation}
  \tilde{\rho} = (\sigma_y \otimes \sigma_y) \rho^{\ast}  (\sigma_y \otimes
  \sigma_y), \qquad \sigma_y = \left(\begin{array}{cc}
    0 & - i\\
    i & 0
  \end{array}\right), \qquad R_{\rho} = \sqrt{\sqrt{\rho}  \hspace{0.17em}
  \tilde{\rho} \hspace{0.17em} \sqrt{\rho}} .
\end{equation}
where $\rho^{\ast}$ denotes complex conjugation. Let the eigenvalues of $R_{\rho}$ be
\[ \nu_1 \geq \nu_2 \geq \nu_3 \geq \nu_4 \geq 0. \]
The Wootters concurrence is {\cite{Wootters1998}}
\begin{equation}
  C (\rho) = \max \{0, \nu_1 - \nu_2 - \nu_3 - \nu_4 \} .
  \label{eq:bell-concurrence-definition}
\end{equation}

The negativity is defined by {\cite{VidalWerner2002}}
\begin{equation}
  \mathcal{N} (\rho) = \frac{\| \rho^{T_B} \|_1 - 1}{2} = \sum_{\mu_j < 0} |
  \mu_j |, \label{eq:bell-negativity-definition}
\end{equation}
where $\rho^{T_B}$ is the partial transposition of $\rho$ with respect to subsystem $B$, and $\mu_j$ are the eigenvalues of $\rho^{T_B}$. For two qubits, the PPT
criterion yields {\cite{Peres1996,Horodecki1996}}
\begin{equation}
  \rho \in \mathcal{S} \quad \Longleftrightarrow \quad \rho^{T_B} \geq 0.
  \label{eq:bell-PPT}
\end{equation}

Now consider the Bell states
\begin{equation}
  | \Phi^{\pm} \rangle = \frac{| 00 \rangle \pm | 11 \rangle}{\sqrt{2}}, \qquad
  | \Psi^{\pm} \rangle = \frac{| 01 \rangle \pm | 10 \rangle}{\sqrt{2}},
\end{equation}
and denote their projectors by
\[ P_1 = | \Phi^+ \rangle \langle \Phi^+ |, \quad P_2 = | \Phi^- \rangle \langle \Phi^- |,
   \quad P_3 = | \Psi^+ \rangle \langle \Psi^+ |, \quad P_4 = | \Psi^- \rangle
   \langle \Psi^- | . \]
A Bell-diagonal state is
\begin{equation}
  \rho_{\mathrm{B}} = \sum_{j = 1}^4 p_j P_j, \qquad p_j \geq 0, \quad \sum_j
  p_j = 1, \qquad p_{\max} = \max_j p_j .
\end{equation}
A direct calculation gives
\[ \mathrm{spec} (\rho_{\mathrm{B}}^{T_B}) = \left\{ \frac{1}{2} - p_1,
   \frac{1}{2} - p_2, \frac{1}{2} - p_3, \frac{1}{2} - p_4 \right\} . \]
By the PPT criterion, we have
\begin{equation}
  \rho_{\mathrm{B}} \in \mathcal{S} \quad \Longleftrightarrow \quad p_{\max}
  \leq \frac{1}{2} . \label{eq:bell-separability}
\end{equation}
On the other hand, spin flipping leaves each Bell projector invariant, so there is
\[ \tilde{\rho}_{\mathrm{B}} = \rho_{\mathrm{B}}, \qquad R_{\rho_{\mathrm{B}}}
   = \rho_{\mathrm{B}} . \]
It is easy to check that, the concurrence and negativity of $\rho_{\mathrm{B}}$ are
\begin{equation}
  C (\rho_{\mathrm{B}}) = (2 p_{\max} - 1)_+, \qquad \mathcal{N}
  (\rho_{\mathrm{B}}) = \left( p_{\max} - \frac{1}{2} \right)_+,
  \label{eq:bell-CN}
\end{equation}
where $(x)_+=\max\{x,0\}$.

\

\begin{proposition}
  \label{thm:bell-exact-rewritten}Let a two-qubit Bell-diagonal state be
$ \rho_{\mathrm{B}} = \sum_{j = 1}^4 p_j P_j$,
then for every $t>0$, there is
  \begin{equation}
    \mathcal{E}_t (\rho_{\mathrm{B}}) = b_t  \left( p_{\max} -
    \frac{1}{2} \right)_+ = \frac{b_t}{2} C (\rho_{\mathrm{B}}) = b_t\,
    \mathcal{N} (\rho_{\mathrm{B}}) , \label{eq:bell-exact-rewritten}
  \end{equation}
  where
$ b_t = \min \left\{ 1, \frac{\sqrt{2 / 3}}{t} \right\} $.

If $p_1=p_{\max}>\frac12$, denote $s=p_1-\frac12$.
  Then one nearest separable state is
  \begin{equation}
    \sigma_{\ast} = \frac{1}{2} P_1 + \sum_{j = 2}^4 \left( p_j + \frac{s}{3}
    \right) P_j . \label{eq:bell-css-rewritten}
  \end{equation}
  If the largest weight occurs at another position, simply permute the indices accordingly. If
  $p_{\max}\le\frac12$, the state $\rho_{\mathrm{B}}$
  is itself separable, and one may take
  $ \sigma_{\ast} = \rho_{\mathrm{B}} $.
\end{proposition}

\begin{proof}
  By the PPT criterion, the separability condition for a two-qubit Bell-diagonal state is $p_{\max}\le
  \frac12$. So we only need to consider the case $p_{\max}>\frac12$.
  Without loss of generality, let
  \begin{equation}
 p_1 = p_{\max} > \frac{1}{2},  
 \qquad  s = p_1 - \frac{1}{2} > 0.
  \end{equation}
  
Now, let us construct a candidate nearest separable state.  
  Consider the Bell-diagonal state
  \begin{equation}
    \sigma_{\ast} = \frac{1}{2} P_1 + \sum_{j = 2}^4 \left( p_j + \frac{s}{3}
    \right) P_j .
  \end{equation}
  It is easy to verify that this is a valid density matrix.
  
  Next we check the separability of $\sigma_{\ast}$. For $j=2,
  3,4$, we have
\begin{equation}
  p_j + \frac{s}{3}  \le \frac{1}{2} - s + \frac{s}{3}  = \frac{1}{2} - \frac{2 s}{3}   < \frac{1}{2} . 
  \end{equation}
Thus the largest Bell weight of $\sigma_{\ast}$ is exactly $\frac12$. By the separability criterion for Bell-
  diagonal states,
we have $
    \sigma_{\ast} \in \mathcal{S}$.

The difference matrix is
  \begin{equation}
    \Delta_{\ast}  = \rho_{\mathrm{B}} - \sigma_{\ast} = \left( p_1 - \frac{1}{2} \right) P_1 - \frac{s}{3}  \sum_{j = 2}^4
    P_j = sP_1 - \frac{s}{3}  (P_2 + P_3 + P_4) . 
  \end{equation}
Therefore, in the Bell basis, we have
  \begin{equation}
    \Delta_{\ast} = \mathrm{diag} \left( s, - \frac{s}{3}, - \frac{s}{3}, -
    \frac{s}{3} \right) .
  \end{equation}
Obviously, there is $\mathrm{Tr}\Delta_{\ast}=0$.
  
Considering the symmetry, the corresponding optimal element may be chosen as
  \begin{equation}
    x = \mathrm{diag}(3 z, - z, - z, - z), \qquad z \ge 0.
  \end{equation}
Thus, we have
\begin{equation}
    \mathrm{diamspec} (x) = 3 z - (- z)  = 4 z. 
  \end{equation}
So the constraint $\mathrm{diam}\mathrm{spec}(x)\le1$ is equivalent to
  \begin{equation}
    z \le \frac{1}{4} .
  \end{equation}
  
We also have
\begin{equation}
    \|x\|_{\mathrm{HS}}^2  = (3 z)^2 + 3 (- z)^2 = 9 z^2 + 3 z^2 = 12 z^2 . 
  \end{equation}
Hence, there is $\|x\|_{\mathrm{HS}}=2\sqrt3\,z\le r_t=\frac{1}{\sqrt2
  \hspace{0.17em}t}$, or
  \begin{equation}
    z \le \frac{1}{2 \sqrt{6}  \hspace{0.17em} t} .
  \end{equation}
Combining the above two constraints, we have
  \begin{equation}
    z_{\max} = \min \left\{ \frac{1}{4}, \frac{1}{2 \sqrt{6}  \hspace{0.17em}
    t} \right\} .
  \end{equation}
Since $\mathrm{Tr} (\Delta_{\ast} x)
= 4 sz$, there is
\begin{eqnarray*}
    N_t (\Delta_{\ast}) && = \sup_{x \in \mathcal{C}_t}  \mathrm{Tr} (\Delta_{\ast} x)
    = 4 s \min \left\{ \frac{1}{4}, \frac{1}{2 \sqrt{6}  \hspace{0.17em} t}
    \right\} \\
    && = 4 s \min \left\{ \frac{1}{4}, \frac{1}{2 \sqrt{6}  \hspace{0.17em} t}
    \right\}  = s \min \left\{ 1, \frac{2}{\sqrt{6}  \hspace{0.17em} t} \right\} = b_t s.
  \end{eqnarray*}
Since $\sigma_{\ast}\in\mathcal{S}$, this immediately gives the upper bound
  \begin{equation}\label{etrb}
    \mathcal{E}_t (\rho_{\mathrm{B}}) \le b_t s.
  \end{equation}

To prove that $\sigma_{\ast}$ is indeed a nearest separable state,
  we also need to show that for every $\sigma\in\mathcal{S}$, there is
$ N_t  (\rho_{\mathrm{B}} - \sigma) \ge b_t s $.
  Define the Hermitian matrix
  \begin{equation}
    X_t = b_t  \left( P_1 - \frac{1}{4} I_4 \right) .
  \end{equation}
There is
$\mathrm{Tr} X_t = 0$, and the eigenvalues of $X_t$ are
  \[ \frac{3 b_t}{4}, - \frac{b_t}{4}, - \frac{b_t}{4}, - \frac{b_t}{4} . \]
So we have
  \begin{equation}
    \mathrm{diamspec} (X_t)  = \frac{3 b_t}{4} - \left( - \frac{b_t}{4} \right) = b_t \le 1. 
  \end{equation}
On the other hand, there is
\begin{equation}
    \left\| P_1 - \frac{1}{4} I_4 \right\|_{\mathrm{HS}}^2  = \left(
    \frac{3}{4} \right)^2 + 3 \left( \frac{1}{4} \right)^2 = \frac{9}{16} + \frac{3}{16} = \frac{3}{4} . 
  \end{equation}
Thus, we have
  \begin{equation}
    \|X_t \|_{\mathrm{HS}} = b_t  \frac{\sqrt{3}}{2} .
  \end{equation}
  Since $b_t=\min\left\{1,\frac{\sqrt{2/3}}{t}\right\}$, we have in particular $b_t
  \le\frac{\sqrt{2/3}}{t}$. Hence, there is
\begin{equation}
    \|X_t \|_{\mathrm{HS}} \le \frac{\sqrt{3}}{2}  \frac{\sqrt{2 / 3}}{t}  = \frac{1}{\sqrt{2}  \hspace{0.17em} t}  = r_t . 
  \end{equation}
Therefore, $X_t$ is a valid feasible point in the calculation of $N_t$.
  
  Thus, for every separable state $\sigma\in\mathcal{S}$, since $\mathrm{Tr}(\rho_{\mathrm{B}}-\sigma)=0$, we have
\begin{eqnarray}\label{ntrbs}
    N_t  (\rho_{\mathrm{B}} - \sigma) &&\ge \mathrm{Tr} [(\rho_{\mathrm{B}} - \sigma)
    X_t]  = b_t  \mathrm{Tr} \left[
    (\rho_{\mathrm{B}} - \sigma) \left( P_1 - \frac{1}{4} I_4 \right) \right]
    \nonumber\\
    && = b_t  \mathrm{Tr} [(\rho_{\mathrm{B}} - \sigma) P_1]  = b_t  \left[ p_1 - \mathrm{Tr} (P_1 \sigma) \right] . 
  \end{eqnarray}

Any separable state can be written as
\[ \sigma = \sum_i q_i |a_i, b_i \rangle \langle a_i, b_i |, \]
where $q_i\ge0$, $\sum_iq_i=1$.  Since $P_1$
is a maximally entangled Bell state, for any pure product state $|a\rangle\otimes|b\rangle\equiv|a, b \rangle$, there is
  \begin{equation}
    \langle a, b|P_1 |a, b \rangle \le \frac{1}{2} .
  \end{equation}
  Therefore, we have
\begin{equation}
    \mathrm{Tr} (P_1 \sigma) = \sum_i q_i  \langle a_i, b_i |P_1 |a_i, b_i \rangle
     \le \sum_i q_i  \frac{1}{2} = \frac{1}{2} . 
  \end{equation}
Hence, from the result (\ref{ntrbs}), there is
\begin{equation}
    N_t  (\rho_{\mathrm{B}} - \sigma)  \ge b_t  \left( p_1 - \frac{1}{2}
    \right)  = b_t s. 
  \end{equation}
  
  Because this inequality holds for every $\sigma\in\mathcal{S}$, there must be
  \begin{equation}
    \mathcal{E}_t (\rho_{\mathrm{B}}) \ge b_t s.
  \end{equation}
  Combining this with the result (\ref{etrb}) gives
  \begin{equation}
    \mathcal{E}_t (\rho_{\mathrm{B}}) = b_t s = b_t  \left( p_{\max} -
    \frac{1}{2} \right), \qquad p_{\max} > \frac{1}{2} .
  \end{equation}
  Including the separable case, we obtain
  \begin{equation}
    \mathcal{E}_t (\rho_{\mathrm{B}}) = b_t  \left( p_{\max} -
    \frac{1}{2} \right)_+ .
  \end{equation}
\end{proof}

In summary, the spectral geometric entanglement of Bell-diagonal states is related to concurrence and negativity as follows:
\begin{equation}
  \mathcal{E}_t (\rho_{\mathrm{B}}) = b_t  \left( p_{\max} - \frac{1}{2}
  \right)_+ = \frac{b_t}{2} C (\rho_{\mathrm{B}}) = b_t\, \mathcal{N}
  (\rho_{\mathrm{B}}) .
\end{equation}

The above result shows that, for fixed $t$,
the spectral geometric entanglement $\mathcal{E}_t$, concurrence $C$, and
negativity $\mathcal{N}$ yield the same ordering on the family of Bell-diagonal states:
all three depend only on the amount by which the largest Bell weight exceeds $1/2$.
The remaining three weights do not affect the value of the minimum distance, but they do affect the nearest separable state
given by Eq.~\eqref{eq:bell-css-rewritten}.

As mentioned earlier, increasing $t$ may decrease the numerical value of the spectral distance,
thereby reducing the spectral geometric entanglement. This just reflects a change in the metric geometry; it does not imply that the physical entanglement of a fixed quantum state diminishes when the parameter is changed.
For every finite $t>0$, we still have
\[ \mathcal{E}_t (\rho) > 0 \quad \Longleftrightarrow \quad \rho
   \text{ is entangled} . \]

Using Proposition \ref{thm:bell-exact-rewritten}, one can explicitly compute the spectral geometric entanglement for some simple examples.

Example 1: The maximally mixed state is
\[ \rho_{\mathrm{mix}} = \frac{\mathbb{I}_4}{4} = \frac{1}{4}  (P_1 + P_2 + P_3 + P_4)
   . \]
Since $p_{\max}=1/4$,
\[ C (\rho_{\mathrm{mix}}) =\mathcal{N} (\rho_{\mathrm{mix}}) =
   \mathcal{E}_t (\rho_{\mathrm{mix}}) = 0. \]
This example also illustrates that mixtures of entangled pure states can be separable.

Example 2: A pure Bell state.
Take $\rho=P_1=| \Phi^+ \rangle\langle \Phi^+ |$. Here $p_{\max}=1$ and $s=1
/2$, so
\begin{equation}
  C (P_1) = 1, \qquad \mathcal{N} (P_1) = \frac{1}{2},
\end{equation}
and
\begin{equation}
 \mathcal{E}_t (P_1) = \frac{b_t}{2} = \left\{\begin{array}{ll}
    \dfrac{1}{2}, & 0 < t \leq \sqrt{2 / 3},\\[.5em]
    \dfrac{1}{\sqrt{6}  \hspace{0.17em} t}, & t > \sqrt{2 / 3} .
  \end{array}\right. \label{eq:bell-pure-example}
\end{equation}
So when the parameter $t$ in the Dirac operator $D_t$ increases, the value of the spectral geometric entanglement will decrease. When $t\to\infty$, there is $\mathcal{E}_t (P_1)\to 0$.
But as we mentioned above, this just reflects a change in the metric geometry; it does not imply that the physical entanglement disappears.

Example 3: Mixture of two Bell states.
Let
\[ \rho_q = qP_1 + (1 - q) P_2, \qquad 0 \leq q \leq 1. \]
Since
\[ p_{\max} = \max \{q, 1 - q\} = \frac{1}{2} + \left| q - \frac{1}{2}
   \right|, \]
we have
\begin{equation}
C (\rho_q) = |2q - 1|, \quad
    \mathcal{N} (\rho_q) = \left| q - \frac{1}{2} \right|, \quad
    \mathcal{E}_t (\rho_q) = b_t  \left| q - \frac{1}{2} \right| .
\label{eq:bell-two-state-example}
\end{equation}
This family of states is separable only when $q=1/2$.

\section{Discussions and conclusions}\label{sec9}

In this paper, we study the Connes spectral distance as a quantum distance measure and its applications in discriminating quantum states and calculating quantum entanglement. Based on finite matrix algebra, we first construct spectral triples for quantum states. These spectral triples are real and even with KO-dimension 6. In order to describe quantum states, we construct three types of Dirac operator corresponding to the quantum trace distance, Hilbert-Schmidt distance and a kind of generalized spectral distance, respectively. The corresponding Connes spectral distances are invariant under unitary transformations.

We obtain some formulas of the spectral distance.
We find that in there spectral triples, the optimal elements can always be chosen to commute with the difference matrix $\Delta$. This means the optimal elements and $\Delta$ can be simultaneously diagonalized.
This result can greatly simplify the calculations of Connes spectral distances.
We find that there are three different parameter regimes for Connes spectral distances. These parameter regimes also depend on the spectrum of difference matrix.

We derive the explicit formula of spectral distance between two arbitrary pure states. We find that for pure states, the spectral distance is similar to quantum trace distance.
As applications, we show how to use spectral distances to distinguish state pairs that have the same quantum trace distance and Hilbert-Schmidt distance.

Based on the spectral distance, we also propose a new definition of geometric measure of entanglement. We calculate the spectral geometric entanglement of Bell-diagonal states. For comparison, we also analysis some other entanglement measures, such as the
concurrence and negativity. We find that for Bell-diagonal states, this spectral geometric entanglement measure has similar properties to the concurrence and negativity.

These results can help us to understand the relationships among Connes spectral distance, quantum trace distance and other quantum distance measures.
These constructions are significant for studies of geometric structures of finite spectral triples. They can help us to study how to describe quantum states in the framework of noncommutative geometry. These results can also be useful for
the studies of Connes spectral distances in other types of spectral triples.

\section*{Acknowledgments}
This work is partly supported by the
Guangdong Basic and Applied Basic Research Foundation (Grant No.
2024A1515010380), the 2024 Guangdong Province Education Science Planning
Project (Higher Education Special) (No. 2024GXJK455), the 2024 Guangdong
Higher Education Teaching Reform Project.


\begin{thebibliography}{99}
\bibitem{Connes1994} A. Connes, \textit{Noncommutative geometry}
(Academic Press, New York, 1994).

\bibitem{Connes1} A. Connes, ``Compact metric spaces, Fredholm modules and
hyperfiniteness.'' \textit{Ergodic Theory Dynam. Systems}
\textbf{9} 207-220 (1989).

\bibitem{Bimonte} G. Bimonte, F. Lizzi, G. Sparano, ``Distances on a
lattice from non-commutative geometry.'' \textit{Phys. Lett. B}
\textbf{341} 139-146 (1994).

\bibitem{Dai} J. Dai, X.-C. Song, ``Connes' distance of one-dimensional
lattices: general cases.'' \textit{Commun. Theor. Phys.}
\textbf{36} 519 (2001).

\bibitem{Cagnache} E. Cagnache, F. D'Andrea, P. Martinetti, J.-C. Wallet,
``The spectral distance on the Moyal plane.'' \textit{J. Geom.
	Phys.} \textbf{61} 1881-1897 (2011).

\bibitem{Wallet} J. C. Wallet, ``Connes distance by examples: Homothetic
spectral metric spaces.'' \textit{Rev. Math. Phys.}
\textbf{24} 1250027 (2012).

\bibitem{Martinetti} P. Martinetti, L. Tomassini, ``Noncommutative geometry
of the Moyal plane: translation isometries, Connes' distance on coherent
states, Pythagoras equality.'' \textit{Commun. Math. Phys.}
\textbf{323} 107--141 (2013).

\bibitem{DAndrea} F. D'Andrea, F. Lizzi, J. C. Varilly, ``Metric properties
of the fuzzy sphere.'' \textit{Lett. Math. Phys.}
\textbf{103} 183--205 (2013).

\bibitem{Pythagoras} F. D'Andrea, P. Martinetti, ``On Pythagoras theorem
for products of spectral triples.'' \textit{Lett. Math. Phys.}
\textbf{103} 469--492 (2013).

\bibitem{Pythagoras1} F. D'Andrea, ``Pythagoras theorem in noncommutative
geometry'' \textit{Contemporary Mathematics}
\textbf{676} 175-210 (2016).

\bibitem{Scholtz} F. G. Scholtz, B. Chakraborty, ``Spectral triplets,
statistical mechanics and emergent geometry in non-commutative quantum
mechanics.'' \textit{J. Phys. A: Math. Theor.}
\textbf{46} 085204 (2013).

\bibitem{Chaoba} Y. Chaoba Devi, S. Prajapat, A. K. Mukhopadhyay, B.
Chakraborty, F. G. Scholtz, ``Connes distance function on fuzzy sphere and
the connection between geometry and statistics.'' \textit{J. Math.
	Phys.} \textbf{56} 041707 (2015).

\bibitem{Martinettip} P. Martinetti, ``From Monge to Higgs: a survey of distance computations in noncommutative geometry.'' \textit{Contemporary Mathematics.} \textbf{676} 1--46 (2016).

\bibitem{Revisiting} Y. Chaoba Devi, K. Kumar, B. Chakraborty, F. G.
Scholtz, ``Revisiting Connes' finite spectral distance on noncommutative
spaces: Moyal plane and fuzzy sphere.'' \textit{Int. J. Geo. Methods
	Mod. Phys.} \textbf{15} 1850204 (2018).

\bibitem{Kumar} K. Kumar, B. Chakraborty, ``Spectral distances on the
doubled Moyal plane using Dirac eigenspinors.'' \textit{Phys. Rev.
	D} \textbf{97} 086019 (2018).

\bibitem{Chakraborty} A. Chakraborty, B. Chakraborty, ``Spectral distance
on Lorentzian Moyal plane.'' \textit{Int. J. Geo. Methods Mod.
	Phys.} \textbf{17} 2050089 (2020).

\bibitem{Lin1} B. S. Lin, T. H. Heng, ``Connes distance of $2 D$ harmonic
oscillators in quantum phase space.'' \textit{Chin. Phys. B}
\textbf{30} 110203 (2021).

\bibitem{Lin2} B. S. Lin, T. H. Heng, ``Connes spectral distance and
nonlocality of generalized noncommutative phase spaces.''
\textit{Eur. Phys. J. Plus} \textbf{137} 899 (2022).

\bibitem{Clare} P. Clare, C.-K. Li, E. Poon, E. Swartz, ``Noncommutative
distances on graphs: An explicit approach via Birkhoff-James
orthogonality.'' \textit{J. Geom. Phys.} \textbf{213}
105483 (2025).

\bibitem{Kumar1} K. Kumar, ``Calibrated Helstrom geometry on the Bloch ball via Connes spectral distance.'' [arXiv:2606.13824] (2026).

\bibitem{LX} B. S. Lin, Z. H. Xu, J. H. Wang, H. L. Chen,
``Connes spectral distances, quantum discord and coherence of qubits.''
\textit{Phys. Scr.} \textbf{101} 095204 (2026).

\bibitem{Lin4} Z. K. You, B. S. Lin, ``Connes spectral distance on twisted fuzzy torus.'' [arXiv:2608.01110] (2026).

\bibitem{Lin3} B. S. Lin, J. H. Wang, Z. K. You, ``A note on unitary invariance of Connes spectral distances of quantum states.'' [arXiv:2605.13014] (2026).

\bibitem{Nielsen} M. A. Nielsen, I. L. Chuang, \textit{Quantum
	Computation and Quantum Information} (Cambridge University Press,
Cambridge, 2000).


\bibitem{Iochum} B. Iochum, T. Krajewski, P. Martinetti, ``Distances in
finite spaces from noncommutative geometry.'' \textit{J. Geom.
	Phys.} \textbf{37} 100--125 (2001).


\bibitem{Vedral} V. Vedral, M. B. Plenio, M. A. Rippin, P. L. Knight, ``Quantifying entanglement.'' \textit{Phys. Rev. Lett.} \textbf{78} 2275-2279 (1997).

\bibitem{Wootters1998} W. K. Wootters, ``Entanglement of formation of
		an arbitrary state of two qubits.'' \textit{Phys. Rev. Lett.} \textbf{80} 2245-2248 (1998).

\bibitem{VidalWerner2002} G. Vidal, R. F. Werner, ``Computable measure of entanglement.'' \textit{Phys. Rev. A} \textbf{65} 032314 (2002).

\bibitem{Peres1996} A. Peres, ``Separability criterion for density
    matrices.'' \textit{Phys. Rev. Lett.} \textbf{77} 1413-1415 (1996).

    
\bibitem{Horodecki1996} M. Horodecki, P. Horodecki and R. Horodecki,
    ``Separability of mixed states: Necessary and sufficient
    conditions.'' \textit{Phys. Lett. A} \textbf{223} 1-8 (1996).

\end{thebibliography}
\end{document}